\documentclass[onecolumn,american,aps,prb]{article}
\pdfoutput=1
\usepackage{graphicx}  
\usepackage{dcolumn}   
\usepackage{bm}        
\usepackage{verbatim}   
\usepackage{caption} 
\usepackage{subcaption}
\usepackage[export]{adjustbox}
\usepackage{amsmath} 
\usepackage{xcolor}
\usepackage{ulem} 
\usepackage[hidelinks=true,colorlinks=false,pdfborder={0 0 0}]{hyperref}
\usepackage{numprint}

\newcommand{\F}[2]{\frac{#1}{#2}}
\newcommand{\D}{\partial}
\newcommand{\FD}[3]{\left(\frac{\D #1}{\D #2}\right)_{#3}}
\newcommand{\SD}[3]{\left(\frac{\D^2 #1}{\D #2^2}\right)_{#3}}
\newcommand{\BE}{\begin{equation}}
\newcommand{\EE}{\end{equation}}

\newcommand{\run}[1]{#1}
\newcommand{\rdeux}[1]{#1}
\newcommand{\edac}[1]{#1}
\newcommand{\rtrois}[1]{#1}

\newcommand{\ardeux}[1]{#1}
\newcommand{\artrois}[1]{#1}

\begin{document}
\title{Numerical simulations of unsteady viscous incompressible flows using general pressure equation}
\author{Adrien Toutant, Université de Perpignan,\\ PROMES-CNRS (UPR 8521), adrien.toutant@univ-perp.fr}
\maketitle
\begin{abstract}
In fluid dynamics, an important problem is linked to the knowledge of the fluid pressure. Recently, another approach to study incompressible fluid flow was suggested. It consists in using a general pressure equation (GPE) derived from compressible Navier-Stokes equation. In this paper, GPE is considered and compared with the Chorin's artificial compressibility method (ACM) \edac{and the Entropically damped artificial compressibility (EDAC) method}. The \edac{three} methods are discretized in a staggered grid system with second order centered schemes in space and a third order Runge-Kutta scheme in time. Three test cases are realized: two-dimensional Taylor-Green vortex flow, the traveling wave and the doubly periodic shear layers. It is demonstrated that GPE is accurate and efficient to capture the correct behavior for unsteady incompressible flows. The numerical results obtained by GPE are in excellent agreement with those obtained by ACM\edac{, EDAC} and a classical finite volume method with a Poisson equation. Furthermore, GPE convergence is better than ACM convergence. The proposed general pressure equation (GPE) allows to solve general, time-accurate , incompressible Navier-Stokes flows. Finally, the parametric study realized in terms of Mach and Prandtl numbers shows that the velocity divergence can be limited by an arbitrary maximum and that acoustic waves can be damped.
\end{abstract}

\section{Introduction}
Incompressible Navier-Stokes (INS) equations correspond to a mixture of hyperbolic and elliptic partial differential equations~\cite{landau:1987}. The pressure in~(\ref{INS}) is not an independent thermodynamic variable. It can be seen as a Lagrangian multiplier of the incompressibility constraint. It is determined by the Laplace or Poisson equation:
\BE
\D_\alpha \D_\alpha P=-(\D_\beta u_\alpha)(\D_\alpha u_\beta)\label{poisson}
\EE
The physical meaning of~(\ref{poisson}) is that in a system with infinitely fast sound propagation, any pressure disturbance induced by the flow is instantaneously propagated into the whole domain. The absence of time evolution equation for pressure is a well known difficulty. Indeed, solving the Poisson equation is often the most costly step in simulation. This difficulty has motivated the search of alternative numerical approaches to determine pressure without solving the Poisson equation. Six different ways have been found:
\begin{enumerate} 
\item The so-called artificial compressibility method (ACM) where a pressure evolution equation is postulated~\cite{chorin:1967}. Many modifications of the Chorin method were suggested in order to compute accurate solutions. For example, Ohwada and Asinari~\cite{ohwada:2010} proposed to introduce a dissipation term in order to improve the quality of numerical solution obtained with artificial compressibility method. \run{The addition of the dissipation term of the artificial continuity equation is useful for surpressing checkerboard instability when a collocated grid system is employed. This revisited ACM is noted SCI for surpressing checkerboard instability. In this paper, a staggered grid system is used. It is shown by numerical results that checkerboard instability is avoided thanks to the staggered grid system and without a dissipation term. The dissipation term proposed in this paper allows to obtain the exact pressure evolution in the case of Taylor-Green vortices and to damp accoustic waves.}
\item The Lattice Boltzmann method (LBM) which uses a velocity-space truncation of the Boltzmann equation from the kinetic theory of gases~\cite{benzi:1992}. LBM and ACM have similarities~\cite{ohwada:2011}. A hybrid method between ACM and LBM has been proposed~\cite{asinari:2012}.
\item An inverse kinetic theory which permits the identification of the (Navier-Stokes) dynamical system and of the corresponding evolution operator which advances in time the kinetic distribution function and the related fluid fields~\cite{tessarotto:2008}. The pressure evolution equation obtained by this method is non-asymptotic. The full validity of INS equations is preserved.
\item The kinetically reduced local Navier-Stokes (KRLNS) wich establishes a simplified equation for the grand potential~\cite{ansumali:2005}.
\item \edac{The entropically damped form of artificial compressibility (EDAC) wich employs a thermodynamic constraint to damp pressure oscillations~\cite{clausen:2013}}. 
\item The general pressure equation (GPE) wich derives the exact pressure evolution from compressible Navier-Stokes equation~\cite{toutant:2017}. GPE can be seen as a physical justification of ACM. GPE contains an additional dissipation term. \edac{KRLNS and EDAC contain exactly the same diffusion term. For KRLNS,  the diffusion term is applied to grand potential instead of pressure.} In fact, the KRLNS is exactly the same as GPE if one replaces the negative of the grand potential by pressure. \edac{The difference between EDAC and GPE comes from pressure advection.} \run{In SCI, the dissipation term is chosen proportional to pressure and not proportional to the Laplacian of pressure. Similarities and differences between GPE, ACM, EDAC, KRLNS and SCI are detailled in section 2.}
\end{enumerate}

\edac{KRLNS}~\cite{karlin:2006,borok:2007,hashimoto:2013,hashimoto:2015} \edac{and EDAC}~\cite{clausen:2013} \edac{have been applied for the simulation of different viscous incompressible flows: three-dimensional Taylor-Green vortex flow, steady-state cavity flow, two-dimensional Taylor-Green vortex flow and doubly periodic shear layers.} This paper realizes a similar work with the GPE approach. It proposes a discretization of GPE in a staggered grid system with second order centered schemes in space and a third order Runge-Kutta scheme in time. In this paper, GPE~\cite{toutant:2017} is applied for the simulation of different unsteady viscous incompressible flows: two-dimensional Taylor-Green vortex flow, the traveling wave and the doubly periodic shear layers. \edac{The results obtained with GPE are compared with those obtained by ACM, EDAC and a classical finite volume method with a Poisson equation.} They show that the proposed general pressure equation (GPE) allows to solve general, time-accurate, incompressible Navier-Stokes flows.

The outline of the paper is as follows. The computational approach is described in section 2. Sections 3 presents the results obtained for unsteady analytical solutions of viscous incompressible flows: the two-dimensional Taylor-Green vortex flow and the traveling wave. The more complex case of doubly periodic shear layers is studied in section 4.

\section{Computational approach}

The classical incompressible Navier-Stokes (INS) equations consist of the equation for the momentum and the continuity equation:
\BE
\D_tu_\alpha + \D_\beta  (u_\beta  u_\alpha)+\D_\alpha P= \F{1}{Re}\D_\beta  \D_\beta  u_\alpha~~~~~\D_\beta u_\beta=0\label{INS}
\EE
With the general pressure equation (GPE), the continuity equation is replaced by:
\BE
\D_tP+\F{1}{Ma^2}\D_\beta u_\beta= \ardeux{\F{\gamma}{RePr}}\D_\beta\D_\beta P\label{GPE}
\EE

With artificial compressibility methods (ACM), the previous equation is approximated by:
\BE
\D_tP+\F{1}{Ma^2}\D_\beta u_\beta= 0\label{ACM}
\EE
\run{With SCI, a dissipation term is added to the equation of ACM}
\BE
\run{\D_tP+\gamma P +\F{1}{Ma^2}\D_\beta u_\beta= 0}\label{SCI}
\EE
\run{where $\gamma$ is a positive function of space and time~\cite{ohwada:2010}. It is worth noting that Ohwada and Asinari choose a discretization that introduces an other dissipation term than $\gamma P$. This other dissipation term is proportional to the pressure Laplacian. However, it cannot be compared with the pressure Laplacian used in GPE, EDAC or KRLNS. Indeed, in SCI, the pressure Laplacian is multiplied by the cell size: its contribution is negligible when grid independence is reached.}
\edac{With EDAC, pressure equation is written:}
\BE
\edac{\D_tP+u_\beta\D_\beta P+\F{1}{Ma^2}\D_\beta u_\beta= \ardeux{\F{\gamma}{RePr}}\D_\beta\D_\beta P}\label{edac}
\EE
\edac{The KRLNS method is based on a subtitution,}
\BE
\edac{\mathcal{G}=P-E_k}\label{grandpotential}
\EE
\edac{where $\mathcal{G}$ is stated to be the negative of the grand potential and $E_k$ is the kinetic energy. The time-evolution of $\mathcal{G}$ is given by:}
\BE
\edac{\D_t\mathcal{G}+\F{1}{Ma^2}\D_\beta u_\beta= \ardeux{\F{\gamma}{RePr}}\D_\beta\D_\beta \mathcal{G}}\label{krlns}
\EE
ACM (eq.~(\ref{ACM})) corresponds to the limit of GPE (eq.~(\ref{GPE})) with $Pr=+\infty$. The diffusive term in GPE\edac{, KRLNS and EDAC} can be surprising. GPE is derived from a rigorous asymptotic analysis of the compressible Navier-Stokes equations. The diffusive term in GPE comes from the conductive heat transfer of the enthalpy equation. Indeed, the enthalpy equation is written in terms of pressure. The conductive heat transfer is written in terms of pressure using the following equation:
\begin{eqnarray}
\D_\beta\D_\beta T&=&\SD{T}{P}{\rho}(\D_\beta P)(\D_\beta P)+\SD{T}{\rho}{P}(\D_\beta\rho)(\D_\beta\rho)+\nonumber\\
&&\FD{T}{P}{\rho}\D_\beta\D_\beta P+\FD{T}{\rho}{P}\D_\beta\D_\beta\rho
\end{eqnarray}
For sufficiently large time-space scales (see the discussion in~\cite{ansumali:2005}), one can neglect density variation ($\rho$, $\chi_T$ and $\alpha$ are supposed constant and evaluated at equilibrium) and temperature variation (around a globally uniform equilibrium temperature) becomes a function of pressure:
\BE
\D_\beta\D_\beta T\approx\FD{T}{P}{\rho}\D_\beta \D_\beta P=\F{\chi_T}{\alpha_\rho}\D_\beta \D_\beta P
\EE 
where $\chi_T$ is the isothermal compressibility coefficient $\chi_T=\F{1}{\rho}\FD{ \rho}{P}{T}$ and where $\alpha_\rho$ is the isobaric thermal expansion coefficient $\alpha_\rho=-\F{1}{\rho}\left(\F{\D \rho}{\D T}\right)_P$. For more details on the establishment of GPE equation, the reader will refer to~\cite{toutant:2017}. The diffusive term of GPE is the difference from artificial compressibility models. \edac{The difference between GPE and EDAC resides essentially in the advective term of pressure that is neglected in GPE and not in EDAC. As it is shown by the numerical results (see section~\ref{unsteady}), the pressure advection of EDAC conducts to errors for pressure estimation (compared to pressure obtained with INS). \ardeux{However, the pressure advection term is usually neglected in EDAC. In this case, there is therefore no difference between EDAC and GPE. Furthermore,} it is worth noting that these errors will disappear with the physical Mach number that is supposed to be very small. However, in these methods, the Mach number becomes an artificial parameter, which may be larger than the physical Mach number. The other difference between GPE and EDAC comes from the physical interpretation of the diffusive term. For GPE, the diffusive term is physical, it comes from the conductive heat transfer of the enthalpy equation. For EDAC, the diffusive term is an artificially way to minimize density fluctuations. In GPE, the Prandtl number like the Mach number can become an artificial parameter, which may be smaller than the physical Prandtl number. As it is shown by the numerical results (see section~\ref{shear}), a small Prandtl number allows to damp acoustic waves. KRLNS is exactly the same as GPE if one replaces the negative of the grand potential by pressure. In order to clarify the differences between \run{SCI}, EDAC, KRLNS and GPE, one can subtract the different pressure equations. One obtains:}
\begin{subequations}
\begin{eqnarray}
\run{GPE-SCI}&\run{=}&\run{\F{1}{RePr}\D_\beta\D_\beta P+\gamma P}\\
\edac{GPE-KRLNS}&\edac{=}&-u_\beta\D_\beta P+\nonumber\\
&&\edac{\D_\beta (u_\beta E_k)+\F{1}{Re}\D_\beta u_\alpha \D_\beta u_\alpha}\\
\edac{GPE-EDAC}&\edac{=}&-u_\beta\D_\beta P
\end{eqnarray}
\end{subequations}
\edac{It is shown in~\cite{clausen:2013} that} 
\begin{itemize}
\item \edac{the convection of kinetic energy $\D_\beta (u_\beta E_k)$ is likely negligible based on the difference in acoustic and convective speeds.}
\item \edac{the kinetic energy dissipation $\F{1}{Re}\D_\beta u_\alpha \D_\beta u_\alpha$ is important in obtaining a qualitatively correct pressure evolution.}
\end{itemize}
\ardeux{The present numerical results show that pressure advection in EDAC conducts to errors in pressure evolution. However,  the pressure advection term is usually neglected in EDAC. Concerning the Prandtl number, simulations realized with EDAC and KRLNS generally assume $Pr=\gamma$. In this paper, the effect of the ratio $\F{Pr}{\gamma}$ is studied. Because the Prandtl number and the heat capacity ratio only appear in the coefficient of pressure diffusion $\F{\gamma}{Re Pr}$, it is useless to study separately their effect. Consequently, it is assumed without loss of generality that $\gamma=1$ and the effect of the Prandtl number is studied. It is shown that small Prandtl number in GPE allows accoustic wave damping.}

\begin{figure}[h] 
\centering 
\includegraphics[width=0.8\textwidth]{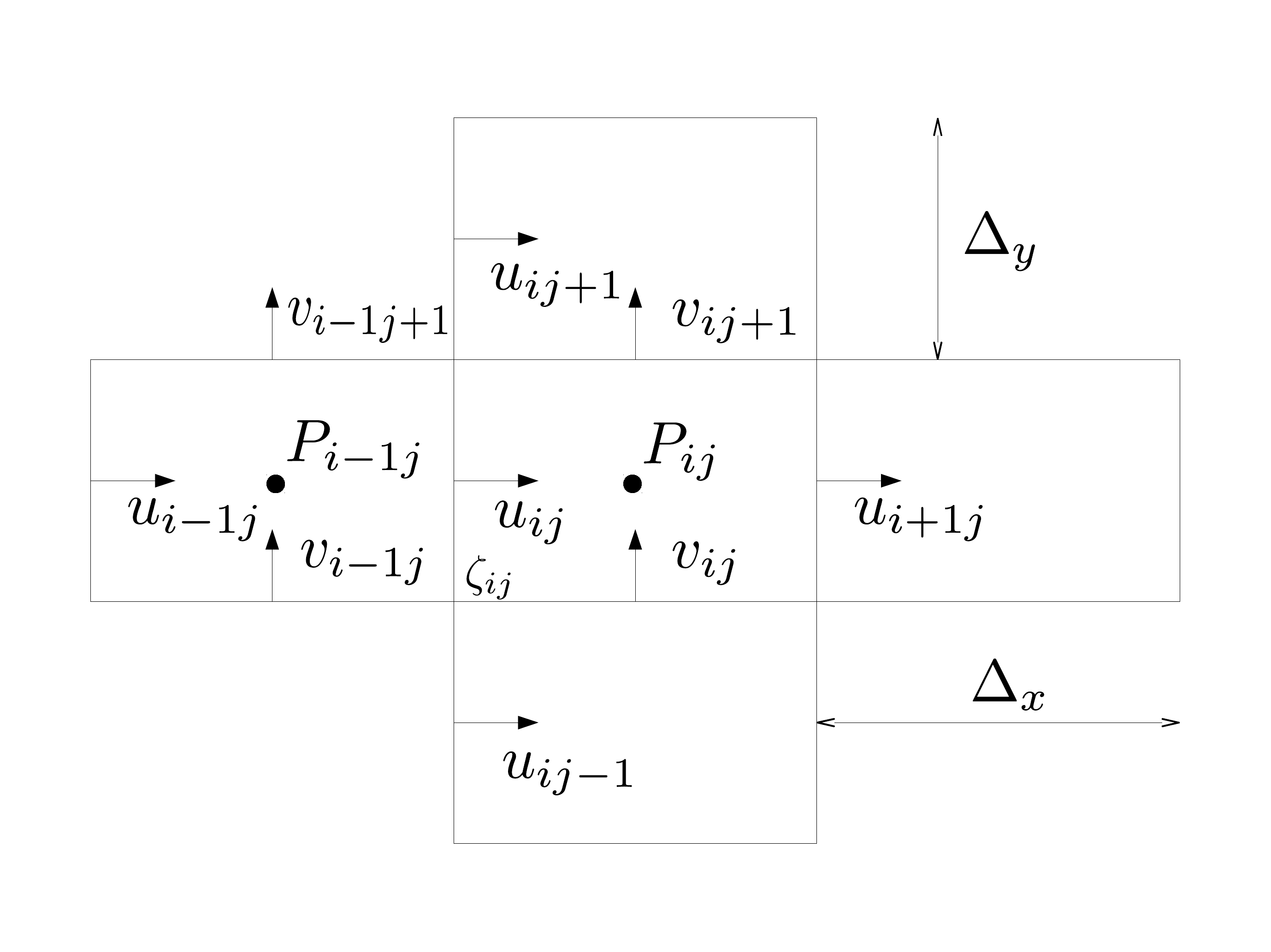}
\caption{Staggered grid system\label{staggered}}
\end{figure}

The equation for the momentum, GPE, ACM and EDAC are discretized with the same scheme without subiteration. The temporal integration is performed using the case 7 of the third order low storage Runge-Kutta schemes proposed by Williamson~\cite{williamson:80}. Due to acoustic, the time scale of INS is related to that of GPE, ACM,\edac{ EDAC} and KRLNS equations; $ t_{GPE} = t_{ACM} = \edac{t_{EDAC} =} t_{KRLNS} = Ma~ t_{INS}$~\cite{karlin:2006}. This means that the time step used in GPE, ACM,\edac{ EDAC} and KRLNS is $Ma$ times smaller than that of INS. \rdeux{This justifies the use of a third Runge-Kutta scheme instead of a fourth Runge-Kutta scheme like in the work of Hashimoto \textit{et al.}}~\cite{hashimoto:2013,hashimoto:2015}. It is worth noting that very low Mach numbers do not imply accuracy problems. They allow very small velocity divergence but they imply very small time-step. One objective of the paper consists in realizing a parametric study on the Mach number to estimate how big the Mach number can be in order to increase the time-step without modifying the results (see section~\ref{param}). Concerning the spatial discretization, we will restrict ourselves to two spatial dimensions to limit the length of the presentation; the extension to 3D is straightforward. The chosen spatial discretization corresponds to a staggered uniform grid system. The velocity components are distributed around the pressure points. The continuity is centered at pressure points. The momentum equation corresponding to each velocity component is centered at the respective velocity point. \rdeux{Staggered grid compared to collocated grid has the advantage that pressure gradient is naturally calculated at velocity points and velocity divergence at pressure points. Furthermore, it allows an easier comparison of the results with INS softwares that use mainly staggered grid system. Finally, staggered grid are generally better for avoiding checkerboard instability. The proposed discretization allows simulations with classical ACM. As it is shown in section~\ref{compaINS}, classical ACM is the most interesting for computational cost because only velocity divergence has to be calculated.} The setting is illustrated in fig.~\ref{staggered}. All used schemes are second order centered schemes:
\begin{subequations}
\begin{eqnarray}
(\nabla\cdot U)_{ij} &\equiv& \F{u_{i+1j}-u_{ij}}{\Delta_x}+\F{v_{ij+1}-v_{ij}}{\Delta_y}\\
(\nabla\cdot (uU))_{ij} &\equiv& 0.25\left(\F{(u_{i+1j}+u_{ij})^2-(u_{i-1j}+u_{ij})^2}{\Delta_x}+\right.\\\nonumber
&&\left.~~~~\F{(u_{ij+1}+u_{ij})(v_{i-1j+1}+v_{ij+1})-(u_{ij-1}+u_{ij})(v_{i-1j}+v_{ij})}{\Delta_y}\right)\\
(\D_x P)_{ij}&\equiv&\F{P_{ij}-P_{i-1j}}{\Delta_x}\\
\edac{(U\cdot\nabla P)_{ij}  }&\equiv& \edac{0.5\left(u_{i+1j}\F{P_{i+1j}-P_{ij}}{\Delta_x}+u_{ij}\F{P_{ij}-P_{i-1j}}{\Delta_x}+\right.}\\\nonumber
&&~~~~\edac{\left.v_{ij+1}\F{P_{ij+1}-P_{ij}}{\Delta_y}+v_{ij}\F{P_{ij}-P_{ij-1}}{\Delta_y}\right)}\\
(\Delta u)_{ij} &\equiv&  \F{u_{i+1j}-2u_{ij}+u_{i-1j}}{\Delta_x^2}+\F{u_{ij+1}-2u_{ij}+u_{ij-1}}{\Delta_y^2}
\end{eqnarray}
\end{subequations}
The convective term $(\nabla\cdot (uU))_{ij}$ is discretized with the standard divergence form in a staggered grid system~\cite{harlow:1965}. It is conservative \textit{a priori} for momentum equation and it is conservative for kinetic energy if $(\nabla\cdot U)_{ij}=0$~\cite{morinishi:1998}. Like LBM, the proposed discretization is local: the pressure and the velocity in one specified grid point only depends of its immediate neighbors of the previous time step. It will greatly facilitate the parallelization of the method and it is anticipated that this method will be very efficient for massively parallel simulation. \rdeux{In particular, using graphics processing unit (GPU) is useful. Hashimoto \textit{et al}}~\cite{hashimoto:2013,hashimoto:2015} \rdeux{realized parallel computations of KRLNS on GPU by using the CUDA library provided by the NVIDIA. They found that the computational time of KRLNS approach using 4th order approximations is 2.7 times faster than that of pseudo spectral method. The speedup of parallel computation is at least 5. Because KRLNS and GPE use the same equation (KRLNS for grand potential and GPE for pressure), the GPE gain of computational time due to parallel computation on GPU is expected to be the same than for KRLNS.}

\section{Unsteady analytical solutions of the Navier-Stokes equation\label{unsteady}}

The analytical solutions of the Navier-Stokes equation simulated in this section are the two-dimensional Taylor-Green vortices and the two-dimensional traveling wave problem. They correspond to unsteady viscous incompressible flows. For GPE and ACM,  they are solved using a $2^n$x$2^n$ computational grid for $n=4$, 5, 6 and 7. The Reynolds, Prandtl and Mach numbers for each run are respectively $Re=100$, $Pr=1$ and $Ma=0.02$ (note that the Prandtl number is not a parameter for ACM). The time step for each run is $\Delta t=10^{-4}$.

\subsection{Decaying vortices}

\begin{figure}[h] 
\centering 
\includegraphics[angle=-90,width=\textwidth]{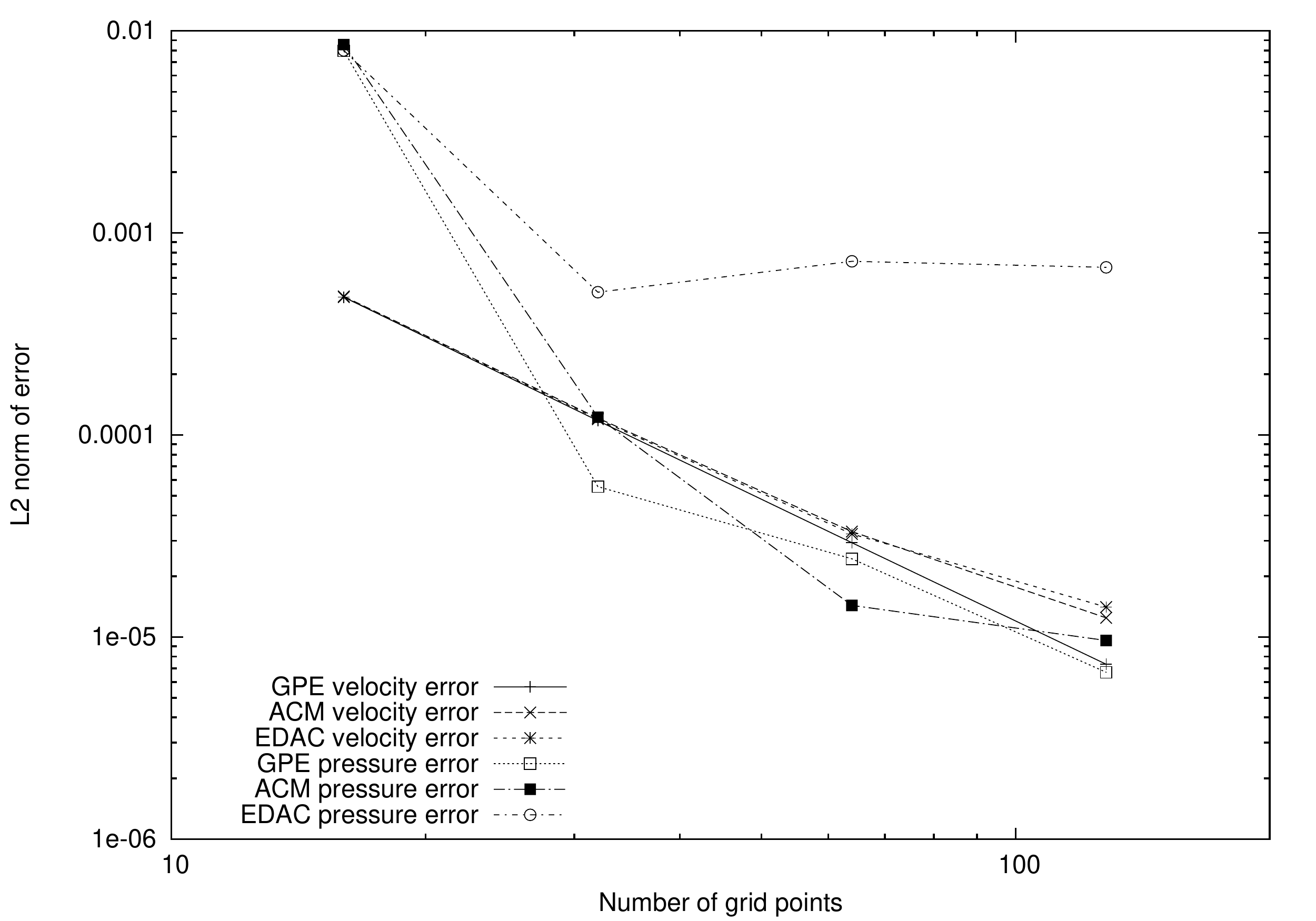}
\caption{\edac{Convergence results for the Taylor-Green vortices. The computed convergence rates are GPE velocity, 2.00; ACM velocity, 1.42; EDAC velocity, 1.2; GPE pressure, 1.87; ACM pressure, 0.57; EDAC pressure, 0.1.}}
\label{compaTaylorGreen}
\end{figure}

The flow considered in this section corresponds to two-dimensional Taylor-Green vortices~\cite{taylor:1937} in a doubly-periodic unit square:
\begin{subequations}
\begin{eqnarray}
u(x,y,t)&=&cos(2\pi x)sin(2\pi y)e^{-\F{8\pi^2}{Re}t}\\
v(x,y,t)&=&-sin(2\pi x)cos(2\pi y)e^{-\F{8\pi^2}{Re}t}\\
p(x,y,t)&=&-\F{1}{4}(cos(4\pi x)+cos(4\pi y))e^{-\F{16\pi^2}{Re}t}
\end{eqnarray}
\end{subequations}
It is worth noting that the previous velocity and pressure fields are exact solutions of INS eq.~(\ref{INS}) and GPE eq.~(\ref{GPE}) with $Pr=1$. They are not solution of ACM eq.~(\ref{ACM}) or KRLNS \edac{or EDAC}. Indeed, one gets~\cite{le:1990}:
\begin{subequations}
\begin{eqnarray}
\D_\beta  (u_\beta  u_\alpha)+\D_\alpha P&=&0\\
\D_tu_\alpha &=& \F{1}{Re}\D_\beta  \D_\beta  u_\alpha\\
\D_\alpha u_\alpha&=&0\\ 
\D_tP&=&\F{1}{RePr}\D_\alpha\D_\alpha P. 
\end{eqnarray}
\end{subequations}
 For each case, the $L_2$ norms of the error in the $u$-component of the velocity and in the pressure were computed at time $t=0.1$ (for symmetry reasons, the $v$-component of the velocity has exactly the same norm of error than the $u$-component). The results are shown in fig.~\ref{compaTaylorGreen}. The convergence rates quoted in the caption of the figure are computed by taking the logarithm to base 2 of the error ratio from 64x64 and 128x128 computations. Because Taylor-Green vortices are exact solution of GPE eq.~(\ref{GPE}) but are not solution of ACM eq.~(\ref{ACM}) \edac{and EDAC eq.~(\ref{edac})}, the convergence rates of GPE are better than those of ACM \edac{and EDAC}. The convergence rate of GPE for the velocity is exactly equal to 2 (the order of the spatial schemes). \edac{For the 128x128 mesh, the L2 norm of GPE error is smaller than for ACM and EDAC.}

\subsection{Traveling wave solution of the Navier-Stokes equations}

\begin{figure}[h] 
\centering 
\includegraphics[angle=-90,width=\textwidth]{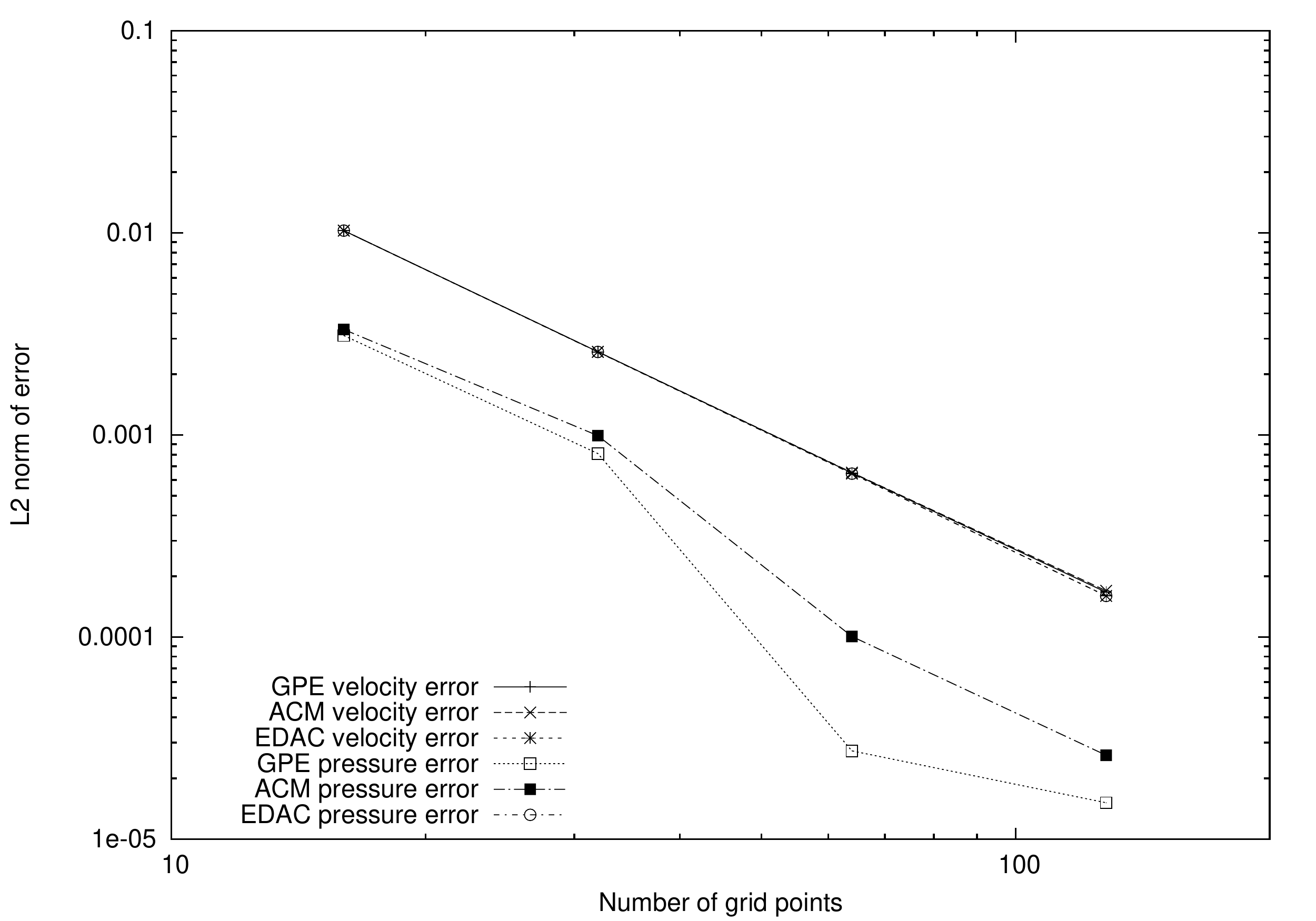}
\caption{\edac{Convergence results for the traveling wave example. The computed convergence rates are GPE velocity, 1.96; ACM velocity, 1.94; EDAC velocity, 2.00; GPE pressure, 0.85; ACM pressure, 1.95; EDAC pressure, 2.00.}}
\label{compawave}
\end{figure}

The flow considered in this section corresponds to the traveling wave solution of the Navier-Stokes equations in a doubly-periodic unit square:
\begin{subequations}
\begin{eqnarray}
u(x,y,t)&=&\F{1}{3}+\F{2}{3}cos(2\pi(x-t))sin(2\pi(y-t))e^{-\F{8\pi^2}{Re}t}\\
v(x,y,t)&=&\F{1}{3}-\F{2}{3}sin(2\pi(x-t))cos(2\pi(y-t))e^{-\F{8\pi^2}{Re}t}\\
p(x,y,t)&=&-\F{1}{9}(cos(4\pi(x-t))+cos(4\pi(y-t)))e^{-\F{16\pi^2}{Re}t}
\end{eqnarray}
\end{subequations}
It is worth noting that the previous velocity and pressure fields are exact solution of INS eq.~(\ref{INS}). They are not solution of GPE eq.~(\ref{GPE}) or ACM eq.~(\ref{ACM}) \edac{or EDAC eq.~(\ref{edac})}. For each case, the $L_2$ norms of the error in the $u$-component of the velocity and in the pressure were computed at time $t=0.7$. The results are shown in fig.~\ref{compawave}. The convergence rates quoted in the caption of the figure are computed by taking the logarithm to base 2 of the error ratio from 64x64 and 128x128 computations. As for the Taylor-Green vortices, \edac{GPE L2 norm of error is smaller than for ACM and EDAC.}

\section{Doubly periodic shear layers\label{shear}}

The case of doubly periodic shear layers is from~\cite{brown:1995,minion:1997}. The initial conditions on the periodic unit square are given by:
\begin{subequations}
\begin{eqnarray}
u&=& \begin{cases}
       tanh(\rho(y-0.25)), & \text{if $y<0.5$}.\\
       tanh(\rho(0.75-y)), & \text{otherwise}.
     \end{cases}\\
v&=& \delta sin(2\pi(x+0.25))
\end{eqnarray}
\end{subequations}
where $\rho$ is the shear layer width parameter and $\delta$ the strength of the initial perturbation. Here, $\rho = 80$, and $\delta = 0.05$ are used. These parameters correspond to those used in~\cite{minion:1997,hashimoto:2013,hashimoto:2015}. The strength of the initial perturbation is the same in~\cite{brown:1995,minion:1997}. The shear layer with parameter of 80 correspond to the thin layer of~\cite{minion:1997} (it is an intermediate value between the thick layer ($\rho = 30$) and the thin layer ($\rho = 100$) from~\cite{brown:1995}). The Reynolds, Prandtl and Mach numbers for each run are respectively $Re=\numprint{10000}$, $Pr=1$ and $Ma=0.02$ (note that the Prandtl number is not a parameter for ACM). The Prandtl and Mach numbers are modified for the parametric study in section~\ref{param}. The time step for each run is $\Delta t=10^{-5}$ (except for the simulations with a Mach number $Ma=0.002$ whose the time step is $\Delta t=10^{-6}$).

\subsection{Mesh independence}

The shear layer problem was computed with 64x64, 128x128, 256x256, 384x384, 512x512, 768x768 and 1024x1024 grids. Mesh independence is studied with velocity, pressure, velocity divergence, kinetic energy, vorticity and enstrophy. Kinetic energy $E_k$, vorticity $\zeta$ and enstrophy $\mathcal{E}$ are defined by 
\begin{subequations}
\begin{eqnarray}
E_k &=&\F{1}{2}(u^2+v^2)\\
\zeta&=&\F{\D v}{\D x}-\F{\D u}{\D y}\\
\mathcal{E}&=&\int_S \zeta^2 dS
\end{eqnarray}
\end{subequations}
Kinetic energy can not be defined uniquely on a staggered grid. In this work, kinetic energy is descretized at pressure points and vorticity is discretized at nodes (see fig.~\ref{staggered}):
\begin{subequations}
\begin{eqnarray}
\left(E_k\right)_{ij} &=&\F{1}{4}(u_{ij}^2+u_{i+1j}^2+v_{ij}^2+v_{ij+1}^2)\\
\zeta_{ij}&=&\F{v_{ij}-v_{i-1j}}{\Delta_x}-\F{u_{ij}-u_{ij-1}}{\Delta_y}
\end{eqnarray}
\end{subequations}

\begin{figure} 
\centering 
\begin{subfigure}{0.45\textwidth}
\centering 
\adjincludegraphics[width=\textwidth]{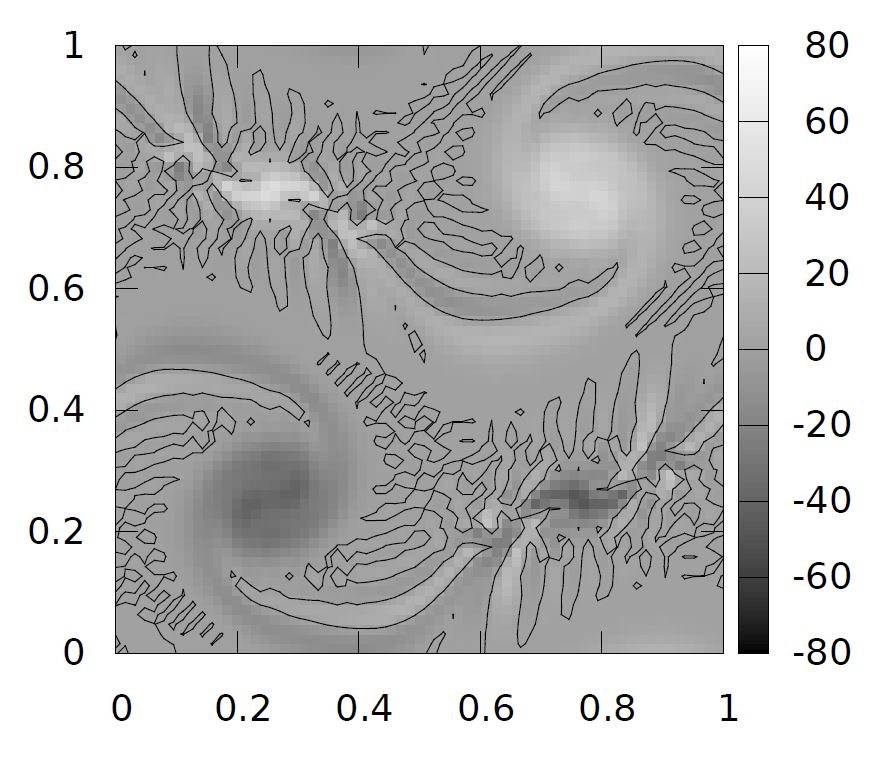}
\caption{Mesh 64x64}
\end{subfigure}
\hspace*{1cm}
\begin{subfigure}{0.45\textwidth}
\centering 
\adjincludegraphics[width=\textwidth]{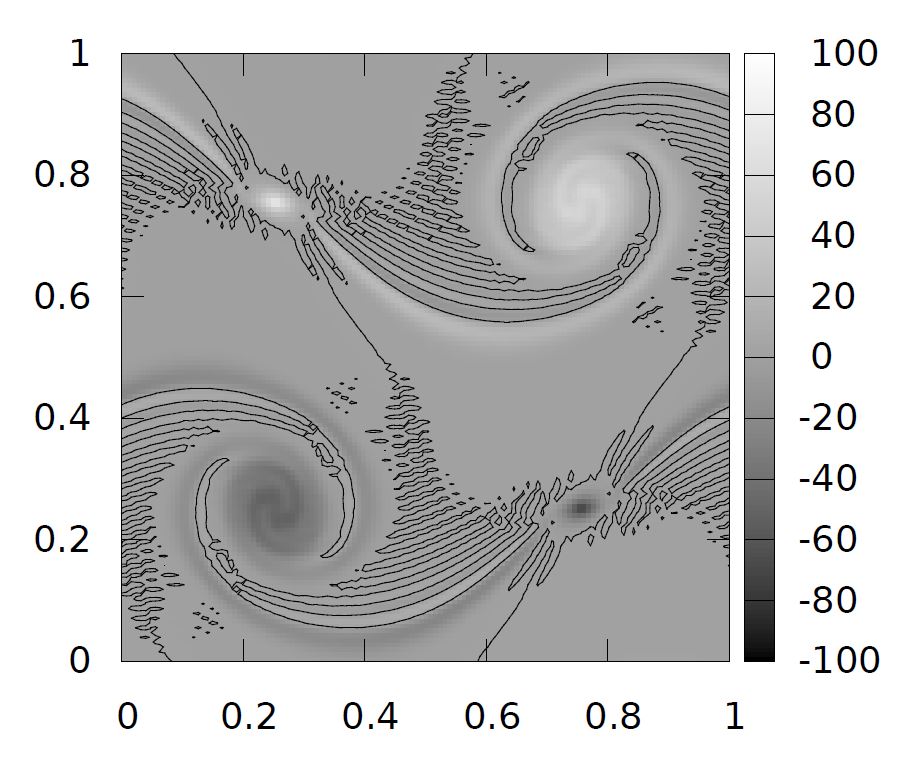}
\caption{Mesh 128x128}
\end{subfigure}

\begin{subfigure}{0.45\textwidth}
\adjincludegraphics[width=\textwidth]{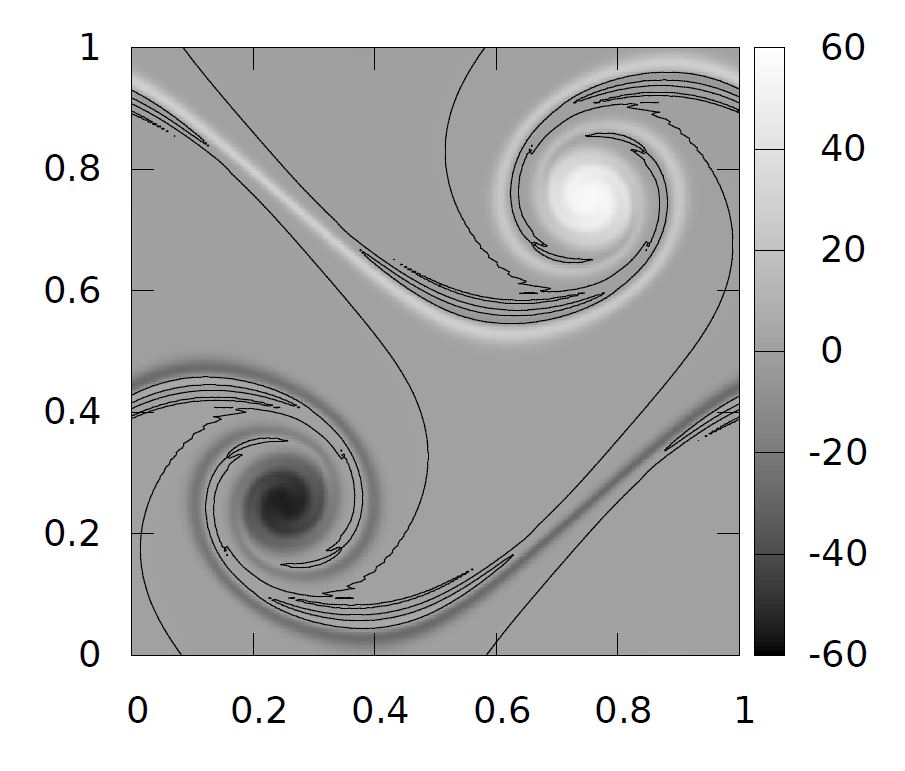}
\caption{Mesh 256x256}
\end{subfigure}
\hspace*{1cm}
\begin{subfigure}{0.45\textwidth}
\adjincludegraphics[width=\textwidth]{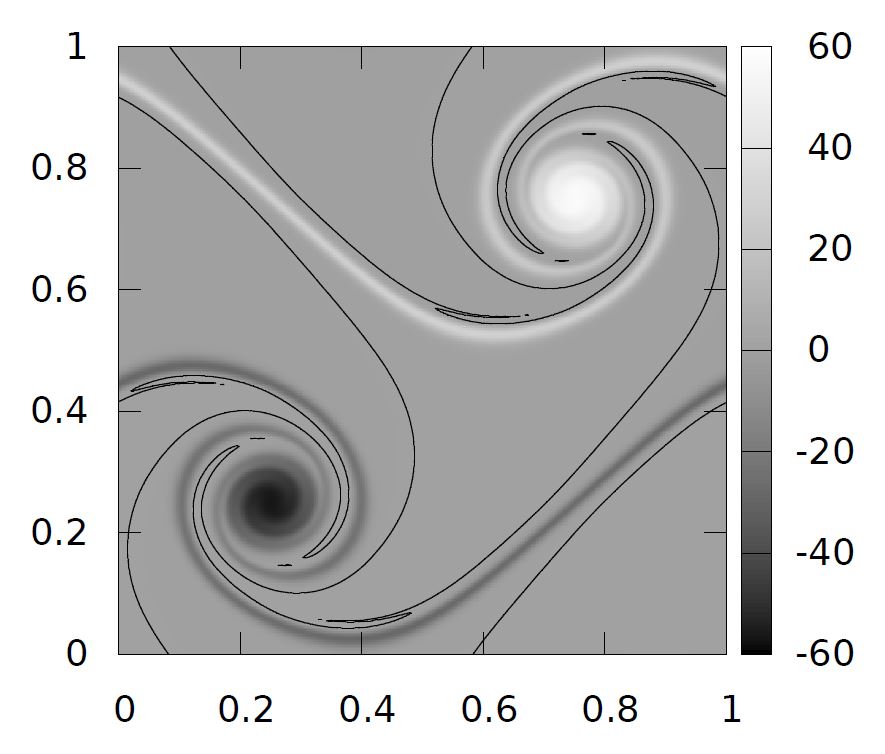}
\caption{Mesh 384x384}
\end{subfigure}

\begin{subfigure}{0.45\textwidth}
\adjincludegraphics[width=\textwidth]{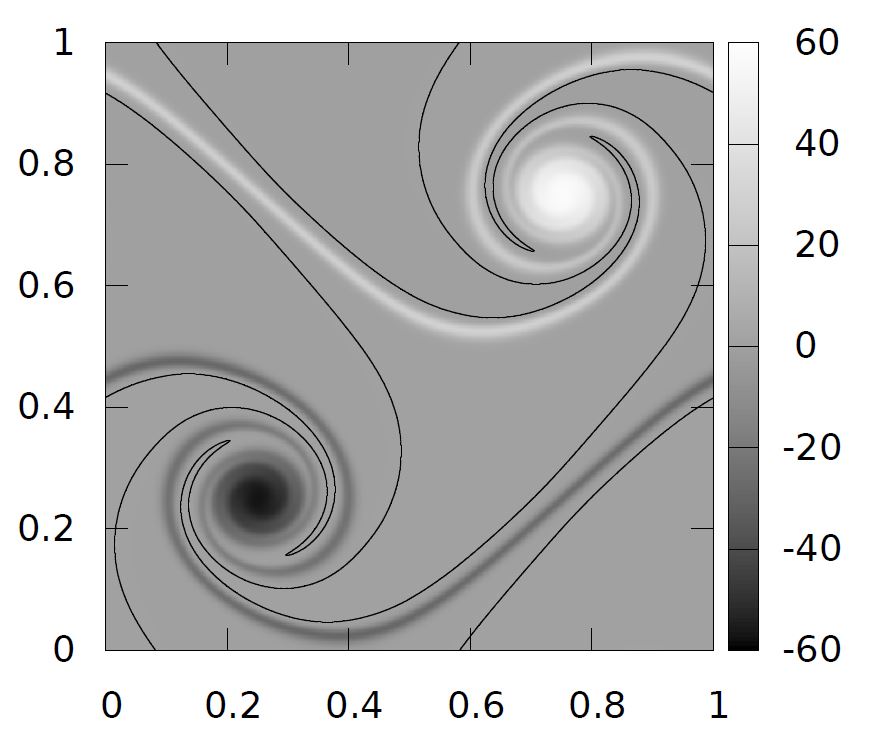}
\caption{Mesh 512x512}
\end{subfigure}
\hspace*{1cm}
\begin{subfigure}{0.45\textwidth}
\adjincludegraphics[width=\textwidth]{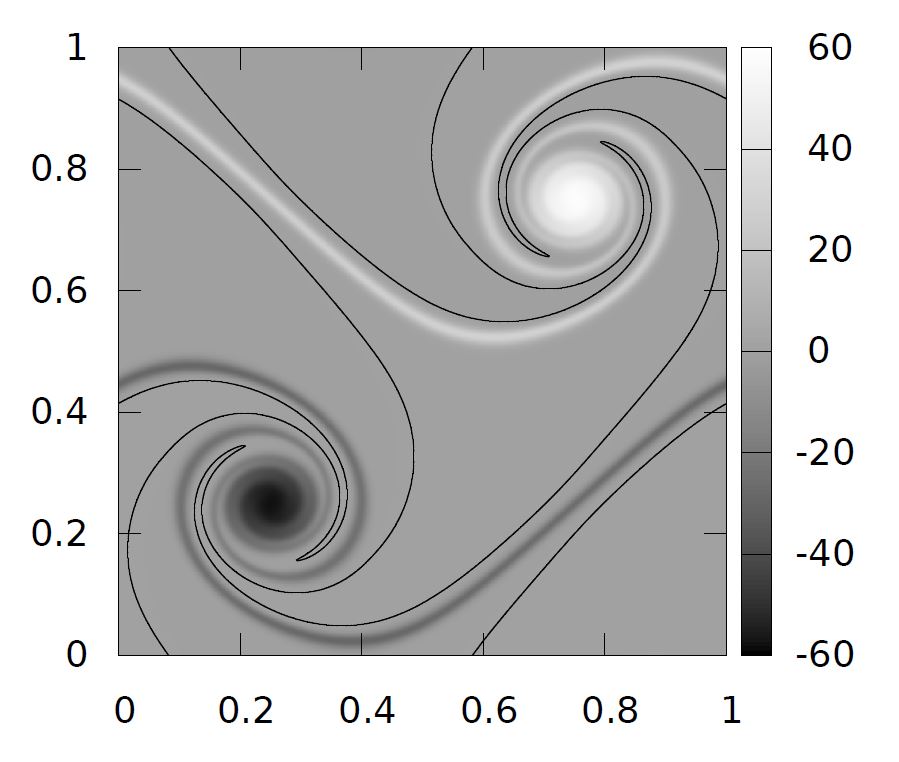}
\caption{Mesh 768x768}
\end{subfigure}

\caption{Vorticity at t=1}
\label{vorticity}
\end{figure}

\begin{figure} 

\begin{subfigure}{0.45\textwidth}
\adjincludegraphics[width=\textwidth]{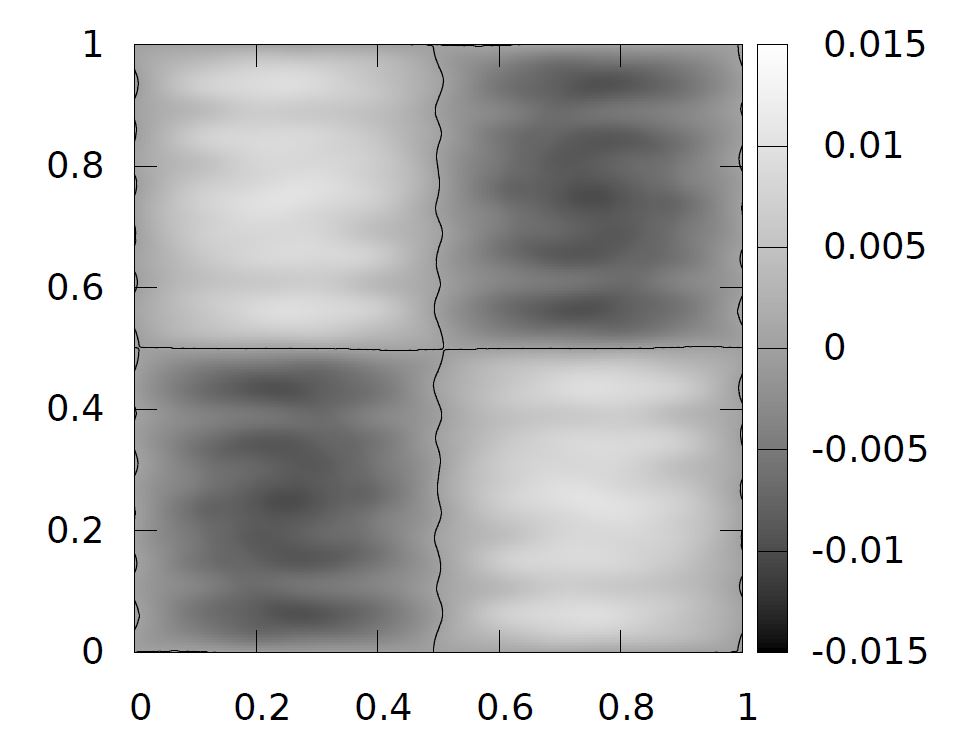}
\caption{Mesh 384x384}
\end{subfigure}
\hspace*{1.5cm}
\begin{subfigure}{0.45\textwidth}
\adjincludegraphics[width=\textwidth]{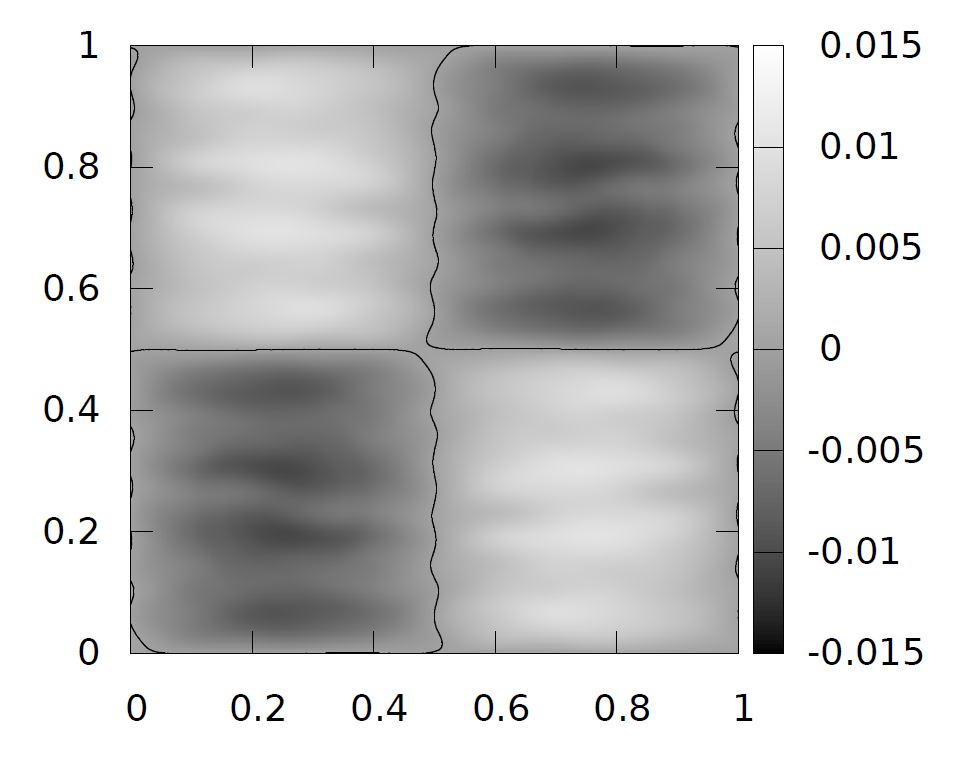}
\caption{Mesh 512x512}
\end{subfigure}

\begin{subfigure}{0.45\textwidth}
\adjincludegraphics[width=\textwidth]{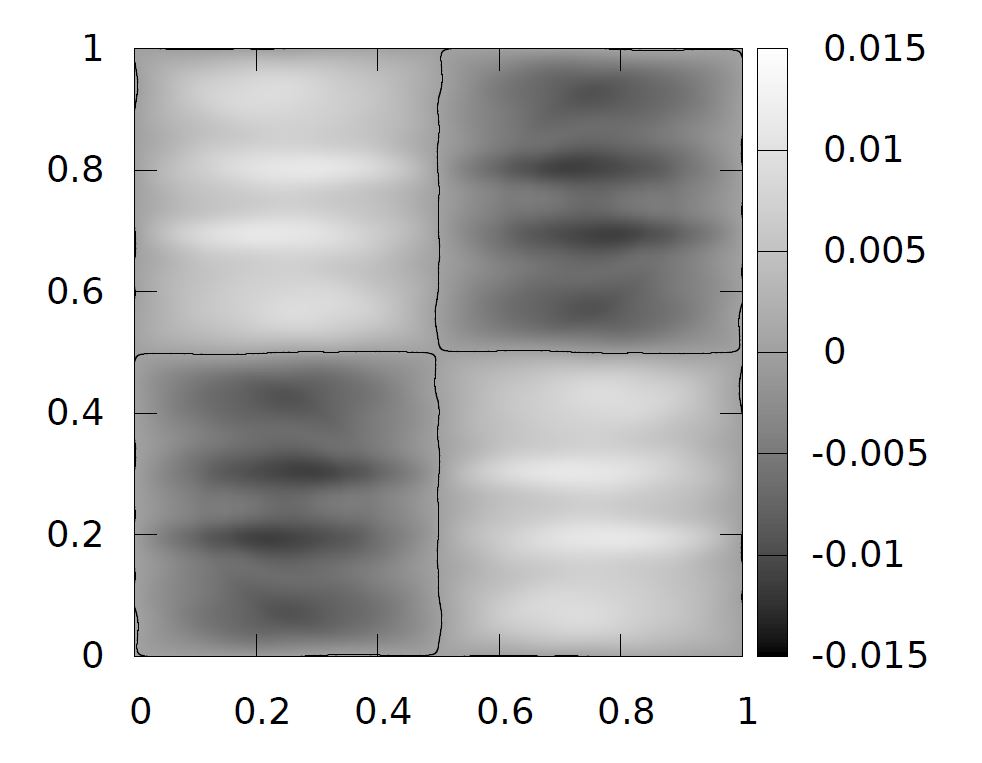}
\caption{Mesh 768x768}
\end{subfigure}
\hspace*{1.5cm}
\begin{subfigure}{0.45\textwidth}
\adjincludegraphics[width=\textwidth]{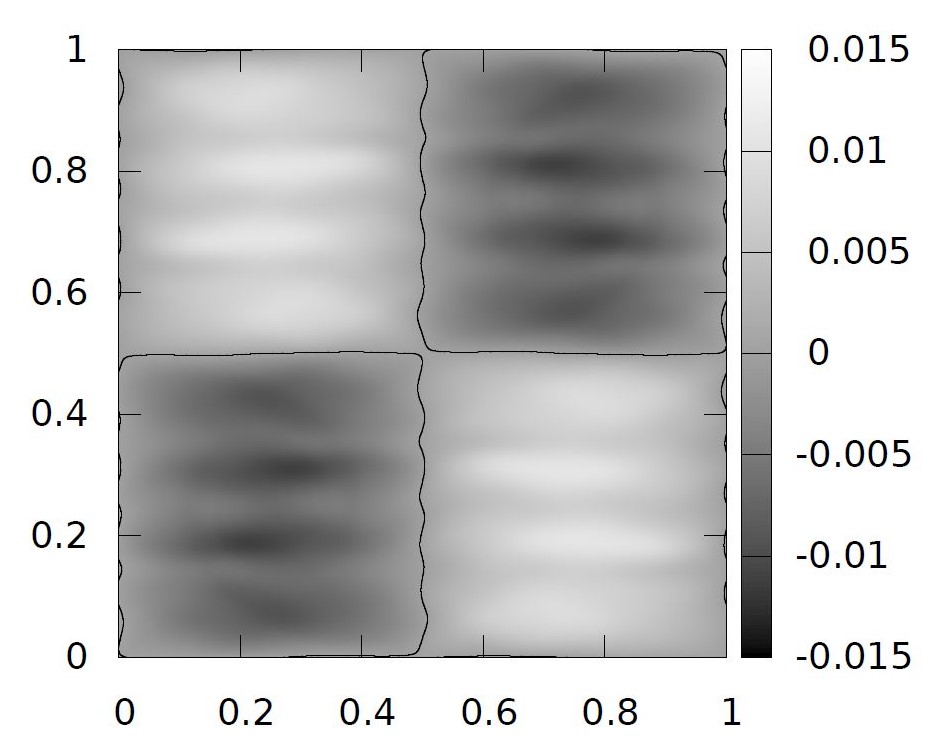}
\caption{Mesh 1024x1024}
\end{subfigure}
\caption{Velocity divergence at t=1}
\label{divergence}
\end{figure}

\begin{figure}
\includegraphics[angle=-90,width=\textwidth]{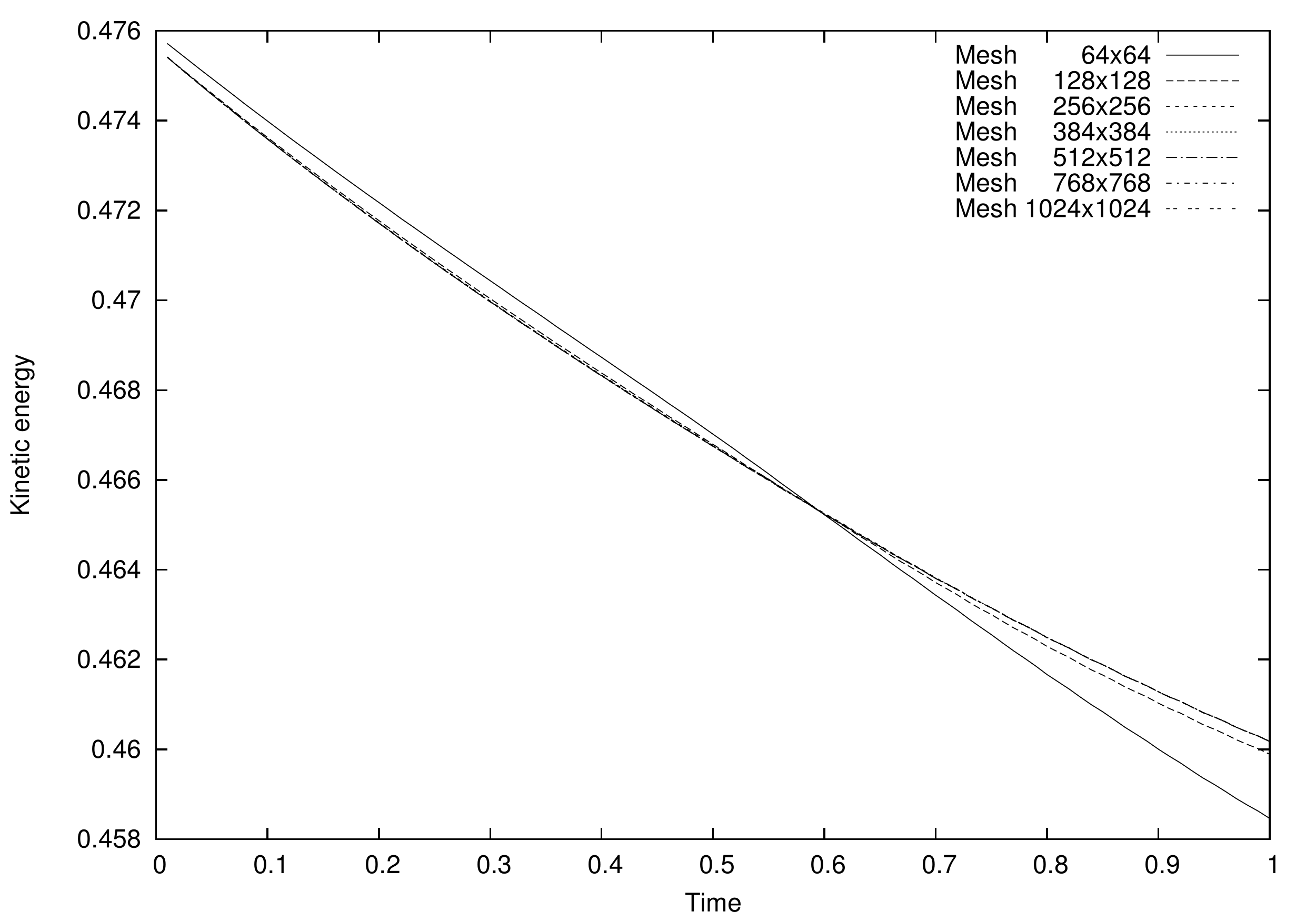}
\caption{Kinetic energy decay}
\label{kdecay}
\end{figure}

\begin{figure}
\includegraphics[angle=-90,width=\textwidth]{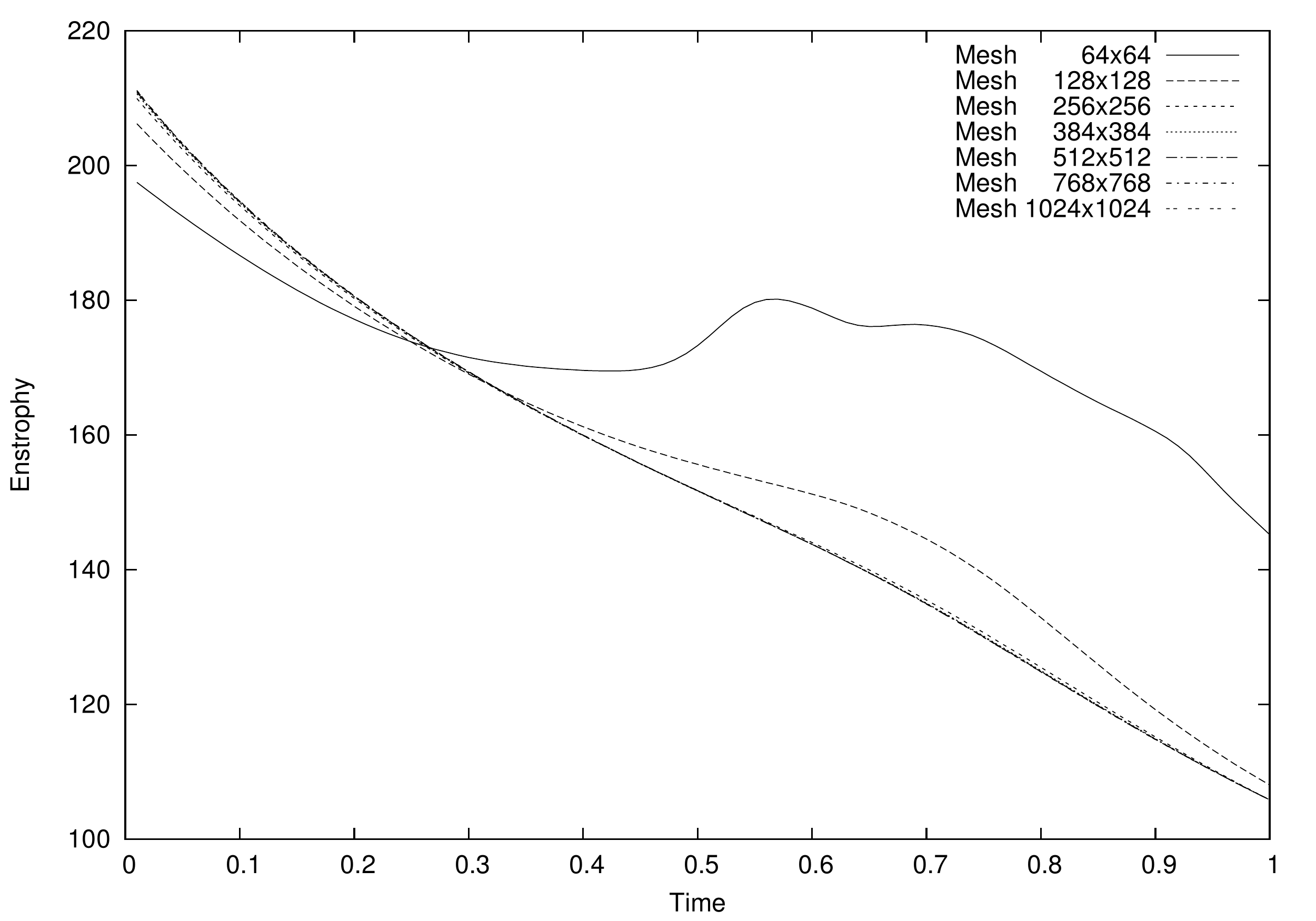}
\caption{Enstrophy decay}
\label{edecay}
\end{figure}

\begin{figure}
\includegraphics[angle=-90,width=\textwidth]{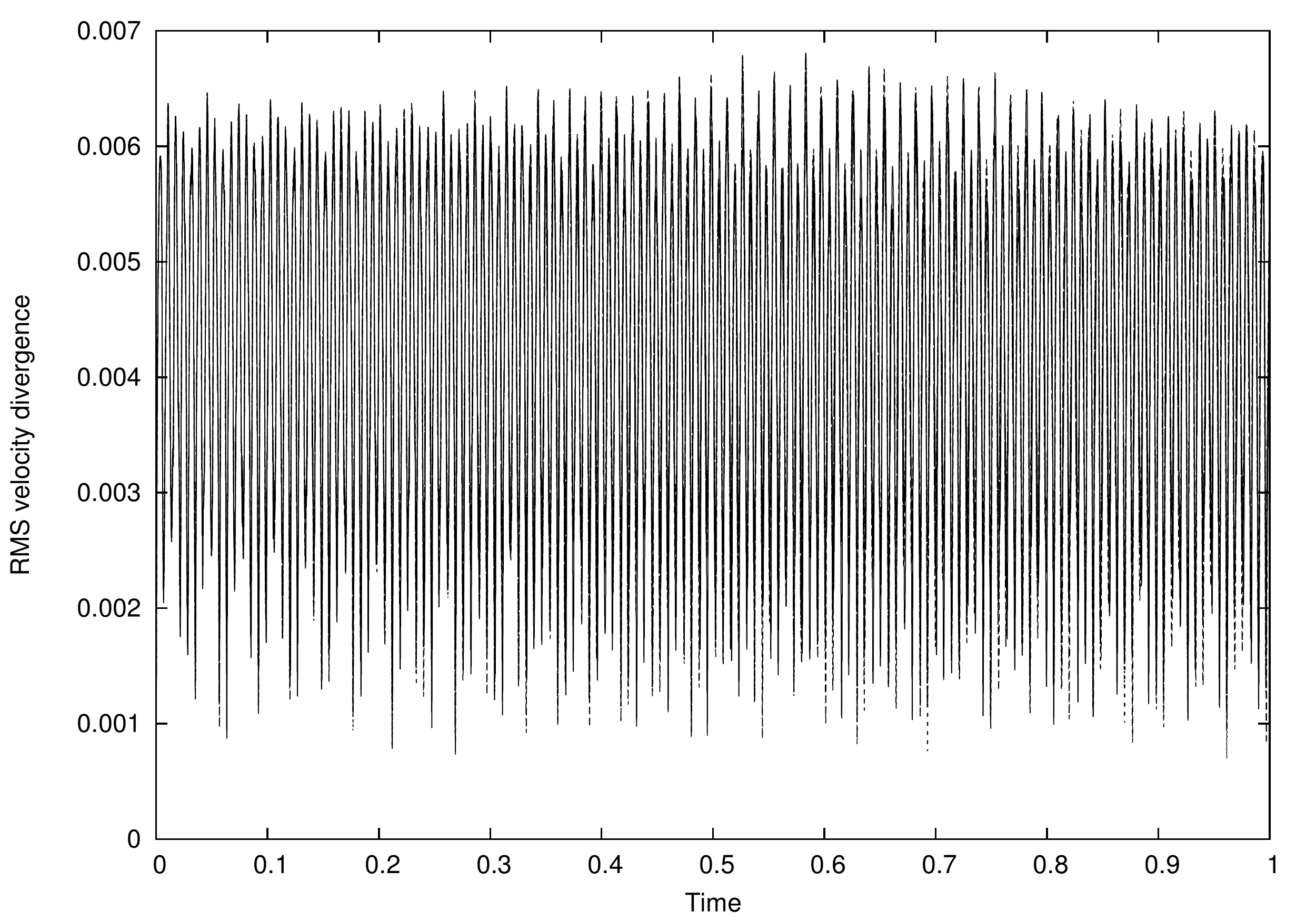}
\includegraphics[angle=-90,width=\textwidth]{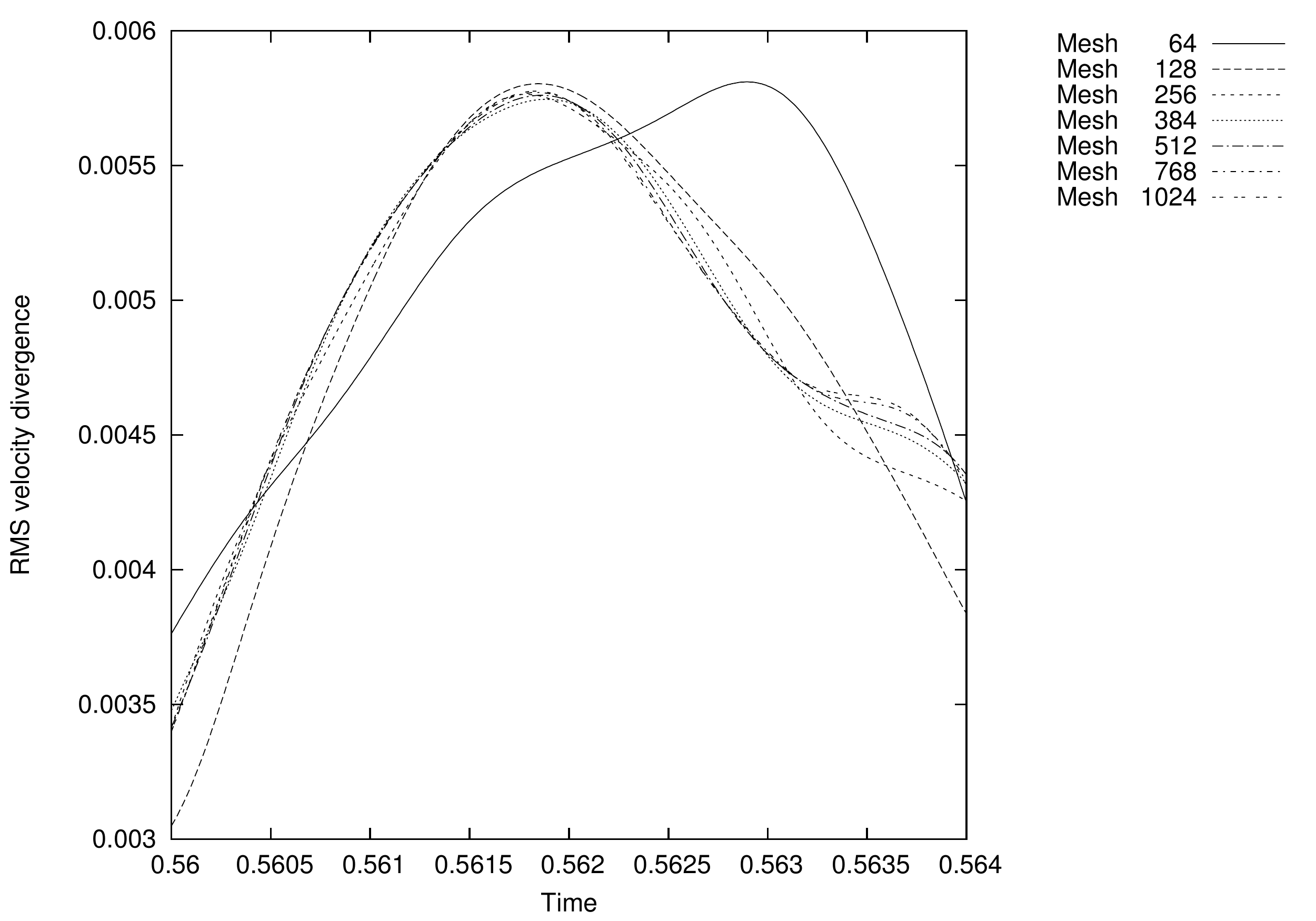}
\caption{Time evolution of root mean square velocity divergence (top: whole time of the simulation, bottom: zoom on one representative peak)}\label{rmsdivergence}
\end{figure}

\begin{figure}
\includegraphics[angle=-90,width=\textwidth]{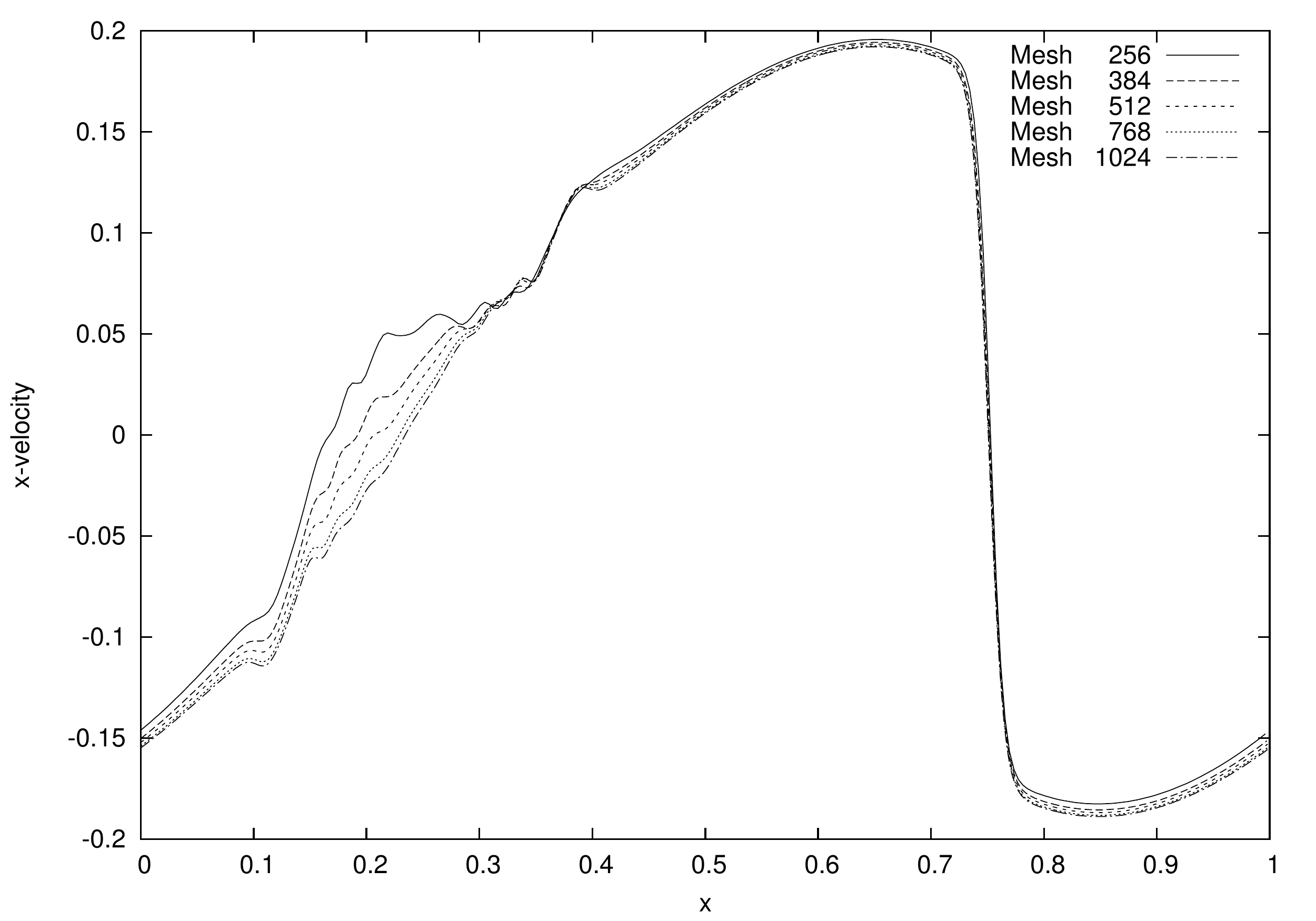}
\caption{Spatial evolution of x-velocity in function of x at y=0.25 and t=1}
\label{vitx}
\end{figure}

\begin{figure}
\includegraphics[angle=-90,width=\textwidth]{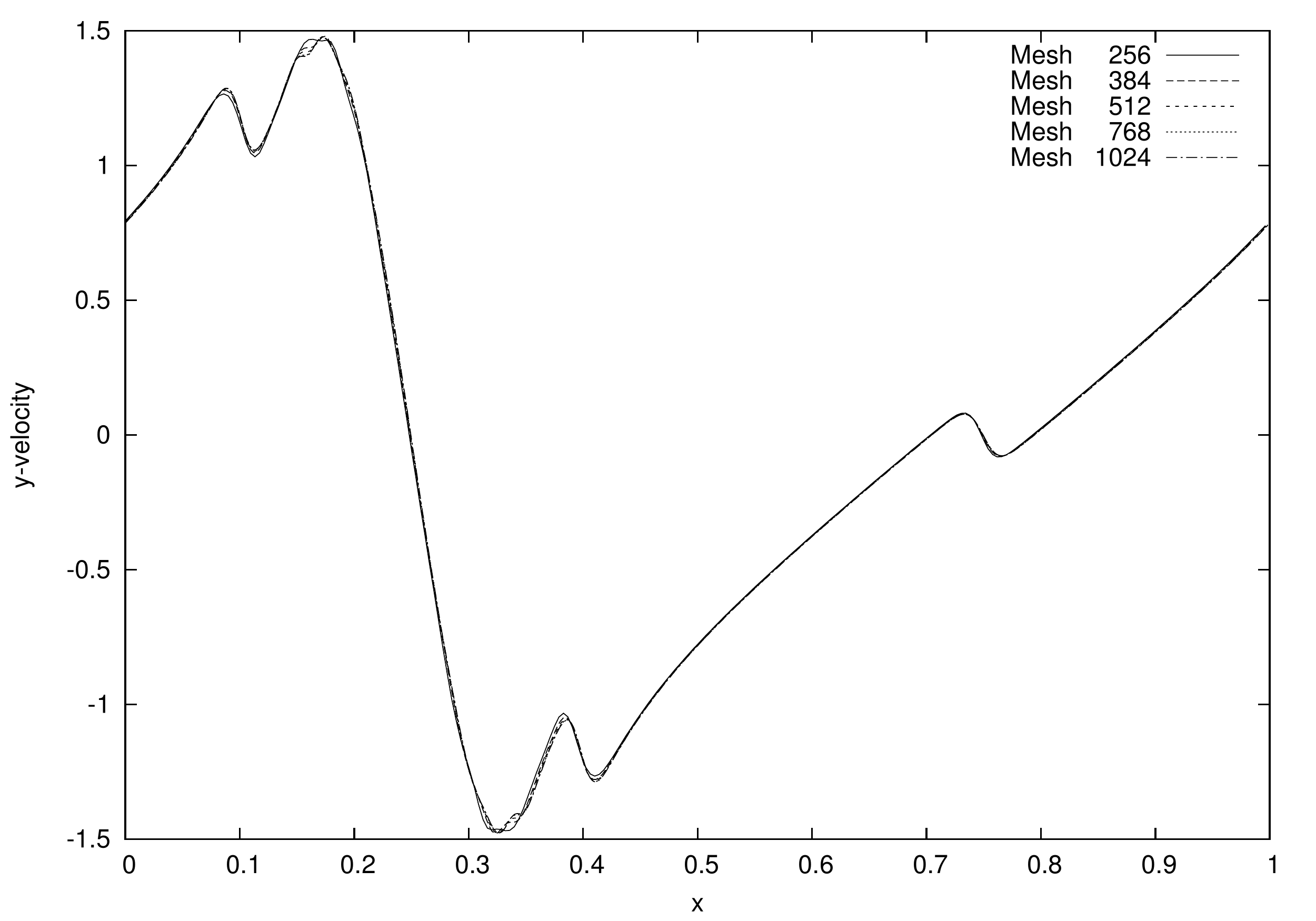}
\caption{Spatial evolution of y-velocity in function of x at y=0.25 and t=1}
\label{vity}
\end{figure}

\begin{figure}
\includegraphics[angle=-90,width=\textwidth]{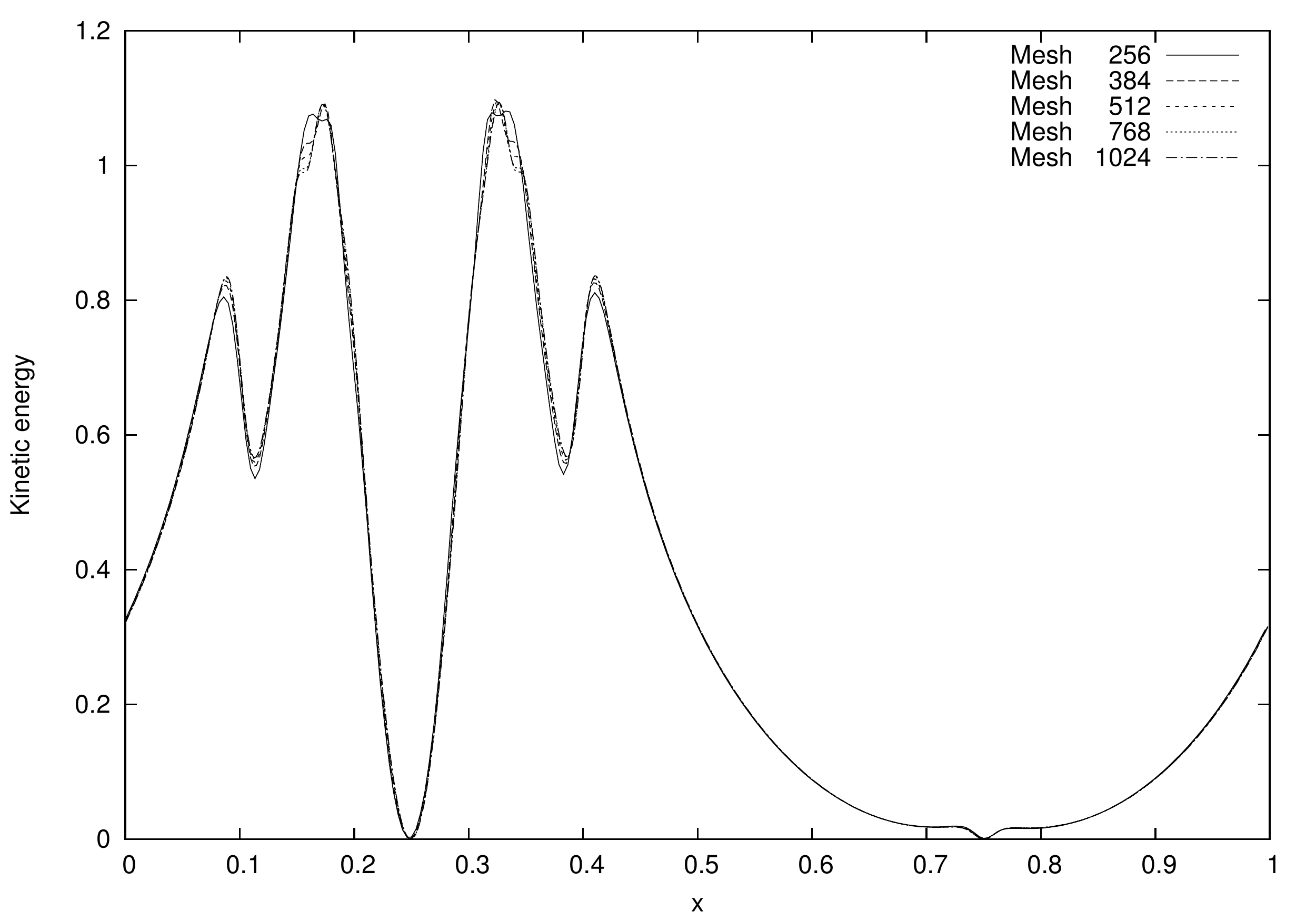}
\caption{Spatial evolution of kinetic energy in function of x at y=0.25 and t=1}
\label{ec}
\end{figure}

\begin{figure}
\includegraphics[angle=-90,width=\textwidth]{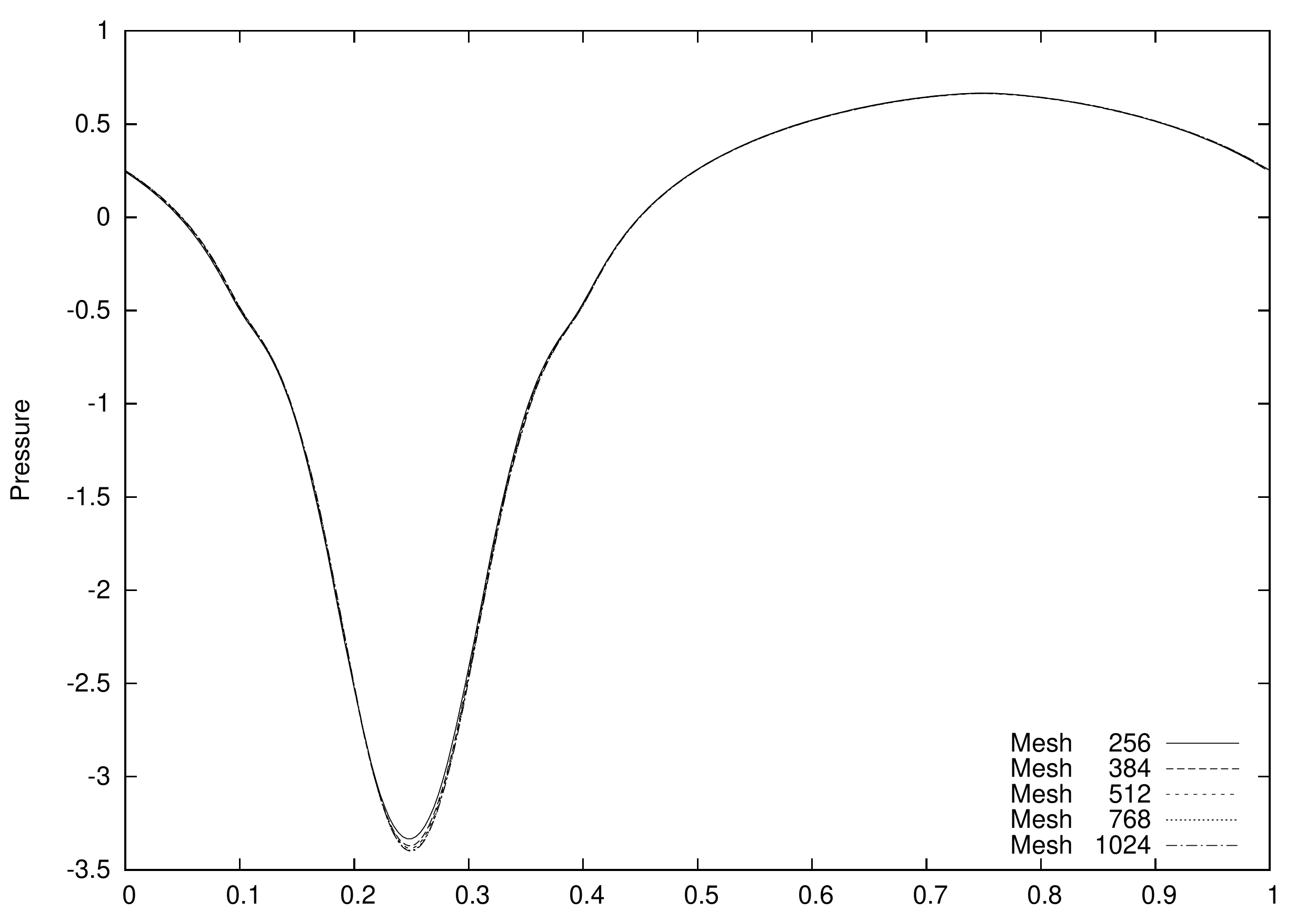}
\caption{Spatial evolution of pressure in function of x at y=0.25 and t=1}
\label{pres}
\end{figure}

\begin{figure}
\includegraphics[angle=-90,width=\textwidth]{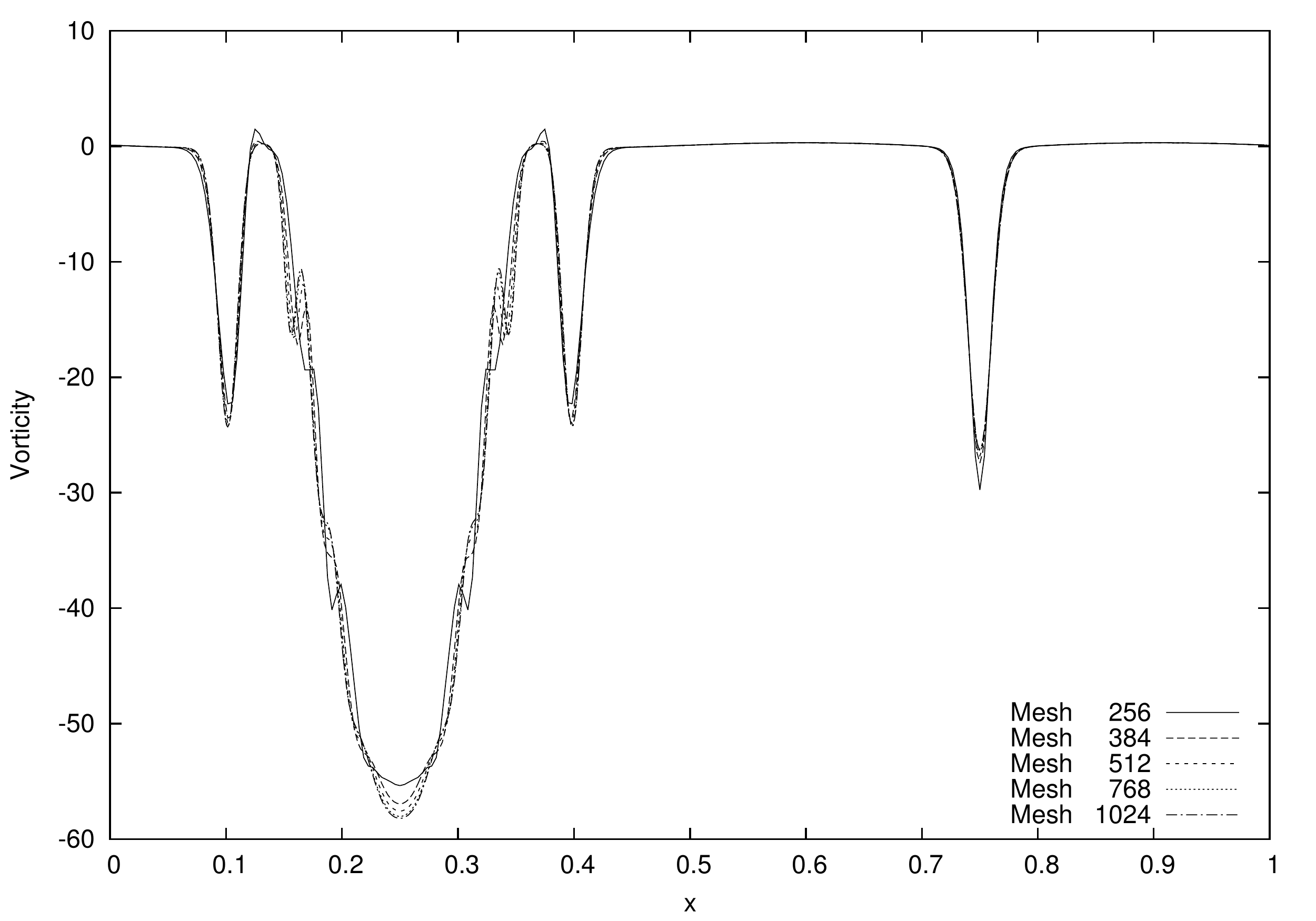}
\caption{Spatial evolution of vorticity in function of x at y=0.25 and t=1}
\label{vort}
\end{figure}

\begin{figure}
\includegraphics[angle=-90,width=\textwidth]{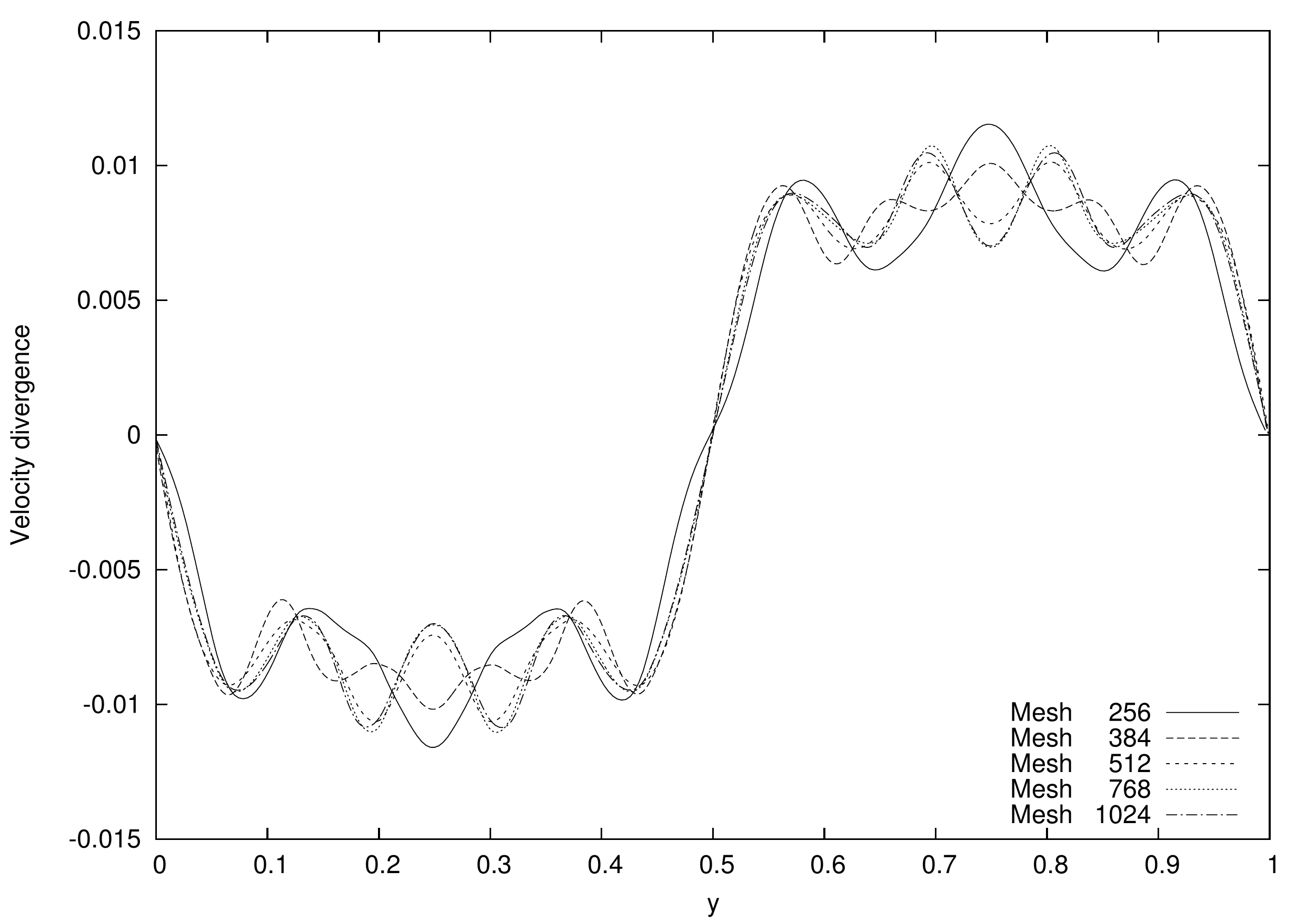}
\caption{Spatial evolution of velocity divergence in function of y at x=0.25 and t=1}
\label{div_xcst}
\end{figure}

Figure~\ref{vorticity} shows the vorticity field at $t=1$ for the different grids. The results from the 64x64 and 128x128 grids show the beginning of unstable oscillations and additional spurious vortices. The additional spurious vortices have disappeared in the 256x256 and 384x384 grids but small oscillations remain. The results from the 512x512 and 768x768 grids appear identical (the 1024x1024 grid -not shown here- gives the same results). It indicates that the 512x512 grid allows computing a reasonably accurate solution. \rtrois{Figure~\ref{divergence} shows the velocity divergence field at $t=1$ for different grids. This figure confirms that the 512x512 grid is sufficient. One can distinguish different frequencies corresponding to the different initial velocity scales. It is worth noting that the average of the velocity divergence is equal to zero}. Figures~\ref{kdecay} and~\ref{edecay} display the computed kinetic energy and enstrophy of the solutions for the different grids. It is clear that these plots are not diagnostic to determine mesh independance (since 256x256 grid seems sufficient with these plots). The time evolution of the root mean square velocity divergence (eq.~(\ref{divurms})) is given in fig.~\ref{rmsdivergence} for the different grids. One notes that the maximum value does not depend on the grid and that again the 512x512 grid is accurate enough. Figures~\ref{vitx} to~\ref{vort} show the spatial evolution in function of x at y=0.25 and t=1 of x-velocity, y-velocity, kinetic energy, pressure and vorticity respectively. Because the velocity divergence evolves faster in the y-direction \rtrois{(see figure~\ref{divergence})}, figure~\ref{div_xcst} represents the spatial evolution in function of y at x=0.25. Note that the point (x,y)=(0.25,0.25) corresponds to the vortex centre. All these figures confirm that the 512x512 grid is accurate enough. This grid is also the one used by~\cite{hashimoto:2013,hashimoto:2015}. Consequently, for the comparison with classical finite volume method with a Poisson equation and for the parametric study, the grid is 512x512.


\clearpage
\subsection{Comparison with incompressible simulations\label{compaINS}}

\begin{figure}
\includegraphics[angle=-90,width=\textwidth]{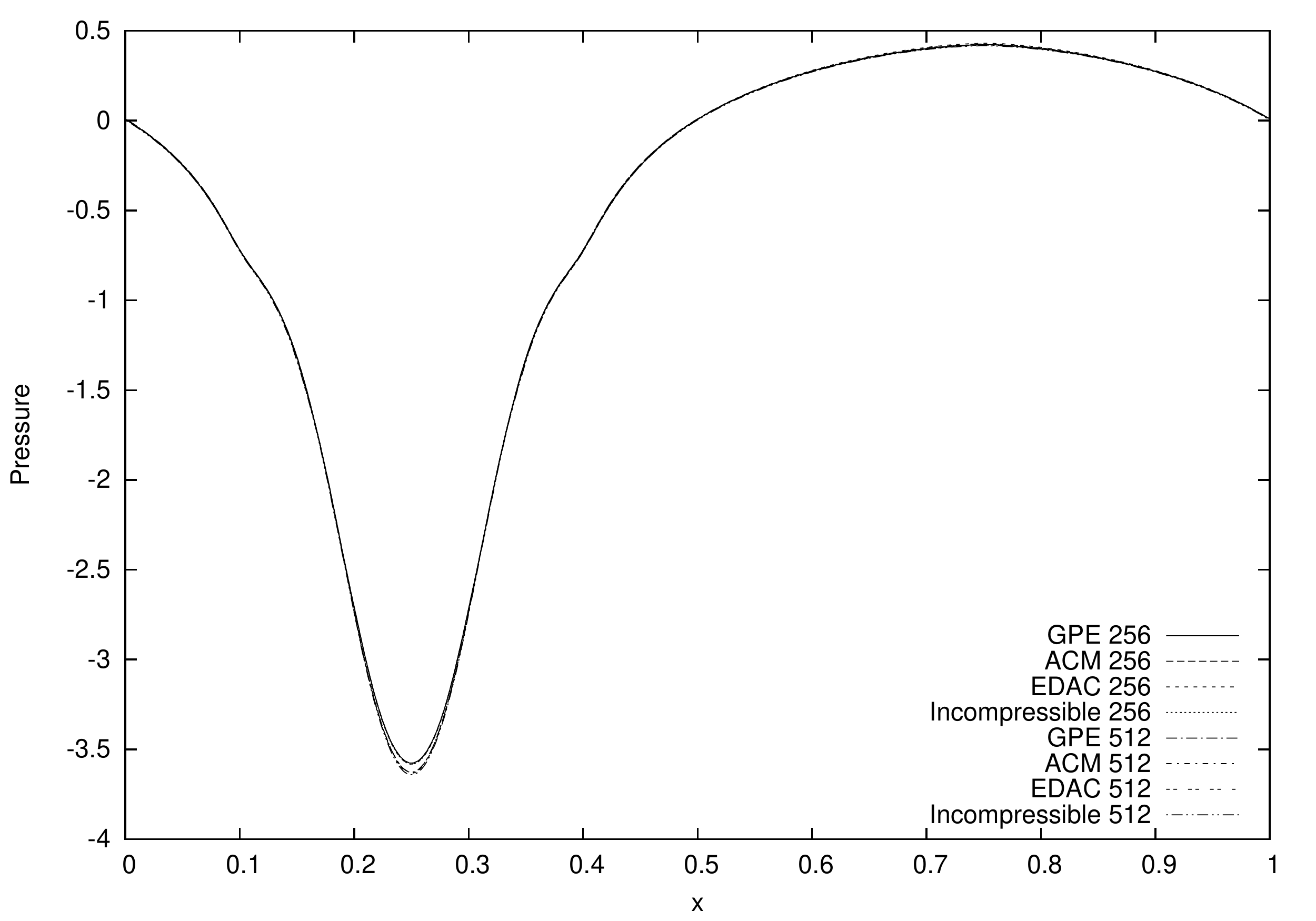}
\includegraphics[angle=-90,width=\textwidth]{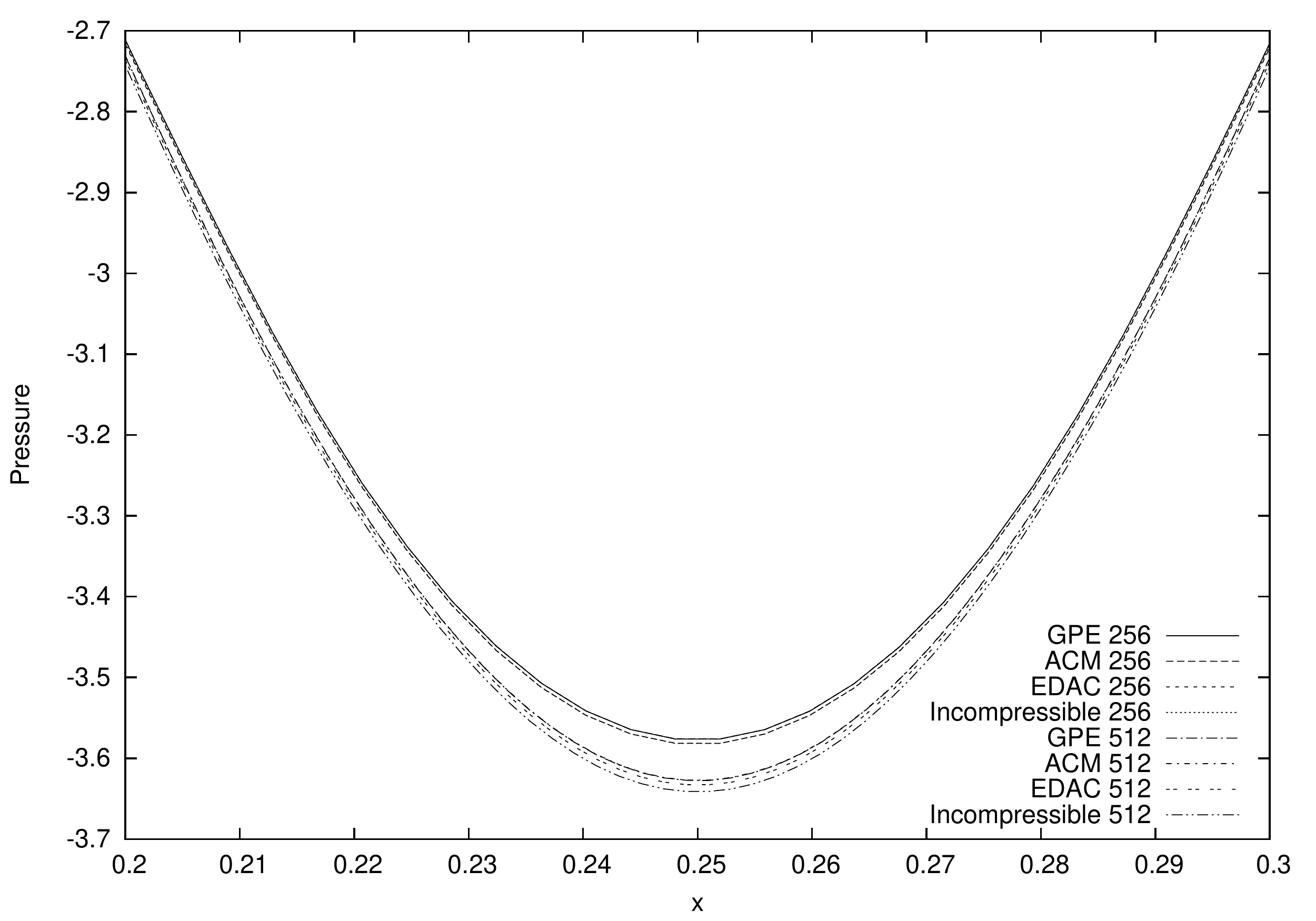}
\caption{Spatial evolution of pressure in function of x at y=0.25 and t=1 (top: whole profile, bottom: zoom around the peak x$\in[0.2,0.3]$). Comparison between incompressible, ACM, \edac{EDAC} and GPE simulations for two different meshes (256x256 and 512x512).}
\label{compap}
\end{figure}

\begin{figure}
\includegraphics[angle=-90,width=\textwidth]{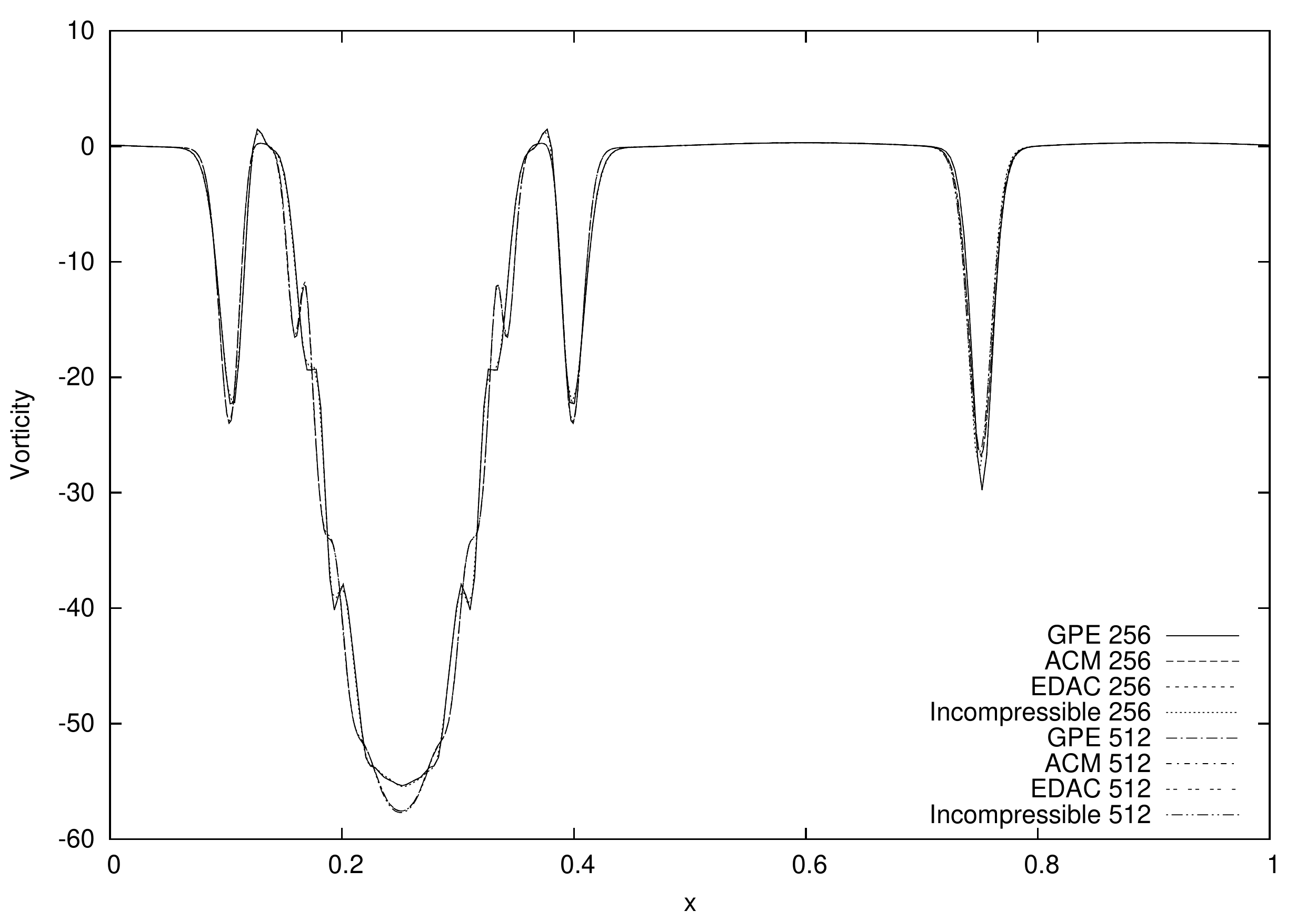}
\includegraphics[angle=-90,width=\textwidth]{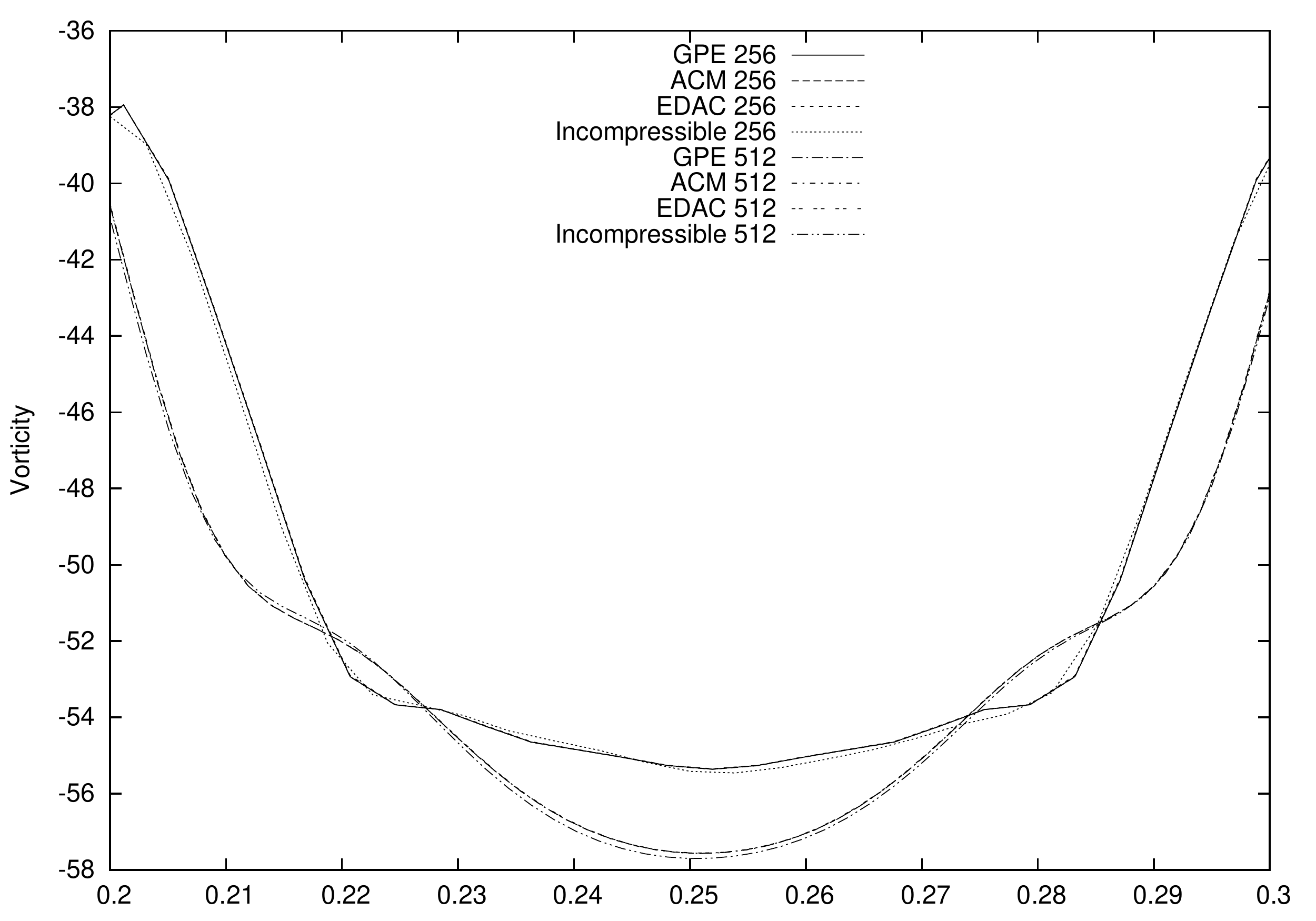}
\caption{Spatial evolution of vorticity in function of x at y=0.25 and t=1 (top: whole profile, bottom: zoom around the peak x$\in[0.2,0.3]$). Comparison between incompressible, ACM, \edac{EDAC} and GPE simulation for two different meshes (256x256 and 512x512).}
\label{compavorticity}
\end{figure}

In this section, the results obtained with GPE, ACM, \edac{EDAC} and INS are compared. The INS simulations are realized with the TrioCFD software developped by CEA (French atomic agency)~\cite{calvin:02}. The selected spatial and temporal schemes in TrioCFD are the same as those implemented for GPE. The INS time step is $\Delta t=10^{-3}$ for the 256x256 grid and $\Delta t=5~10^{-4}$ for the 512x512 grid. The pressure is solved with the classical pressure-Poisson formulation~\cite{chorin:1968,minion:1997}. \rdeux{The Poisson equation is solved with the conjugate gradient method and the Symmetric Successive Over-Relaxation preconditioning technique. With a Mach number of 0.02 for GPE (ACM or EDAC) and a residual norm tolerance of $10^{-12}$ for the Poisson equation (INS), it is found that computational time of GPE, ACM and EDAC is respectively 1.3, 1.5 and 1.2 times smaller than that of INS. ACM is the most interesting in terms of computational time because only velocity divergence has to be calculated. EDAC is the least interesting because pressure advection and diffusion have also to be computed.} Only the pressure and the vorticity fields (fig.~\ref{compap} and~\ref{compavorticity} respectively) are presented  because the results with velocity and kinetic energy are the same. Figures~\ref{compap} and~\ref{compavorticity} indicate that the results obtained with GPE, ACM and INS are in very good agreement. The differences between GPE, ACM \edac{and EDAC} are not distinguishable. The differences with INS are smaller than those due to the mesh. These results demonstrate that the proposed discretization of GPE, ACM \edac{and EDAC} allows describing the time evolution of velocity and pressure without subiteration. \edac{They also show that the additional term in the EDAC equation compared to the GPE equation is negligible. It confirms the asymptotic expansion realized in~\cite{toutant:2017} that conducts to neglect pressure advection ($u_\beta \partial_\beta P$) for small Mach number}.  The test of different time steps for GPE, ACM \edac{and EDAC} shows that $\Delta t=2.5~10^{-5}$ is stable and that $\Delta t=5~10^{-5}$ is not stable for the 512x512 grid. Consequently, it is found that the proposed discretization conducts to the same stability time step for GPE, ACM \edac{and EDAC}. 

\clearpage
\subsection{Prandtl and Mach numbers effect\label{param}}

GPE depends on three dimensionless numbers: the Reynolds, Prandtl and Mach numbers. In this section, a parametric study on the Prandtl and Mach numbers is realized. The effect of the Reynolds number is not studied because it is considered here as a physical parameter while the Prandtl and Mach numbers are considered as numerical parameters. The objective is to estimate 
\begin{itemize}
\item how small the Prandtl number can be in order to damp acoustic waves without modifying the results and,
\item how big the Mach number can be in order to increase the stability time step without modifying the results. 
\end{itemize}
\artrois{The tested Prandtl numbers are 0.001, 0.01, 0.1, 1, 10 and $+\infty$. The tested Mach numbers are 0.2, 0.1, 0.05, 0.025, 0.02 and 0.002.}

\begin{figure}[h]
\includegraphics[angle=-90,width=\textwidth]{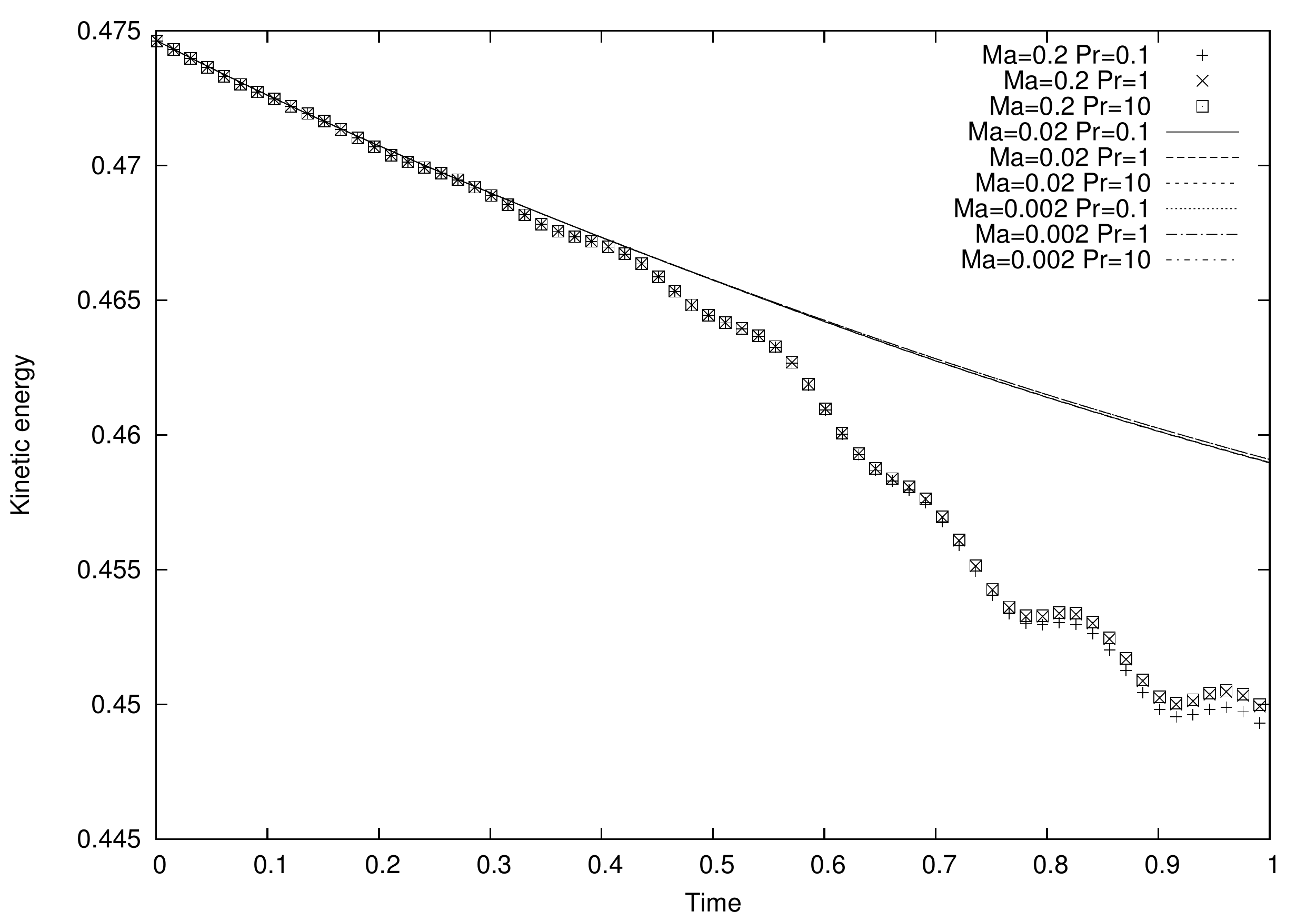}
\caption{Kinetic energy decay for different Mach and Prandtl numbers.}
\label{compakinetic}
\end{figure}

\begin{figure}[h]
\includegraphics[angle=-90,width=\textwidth]{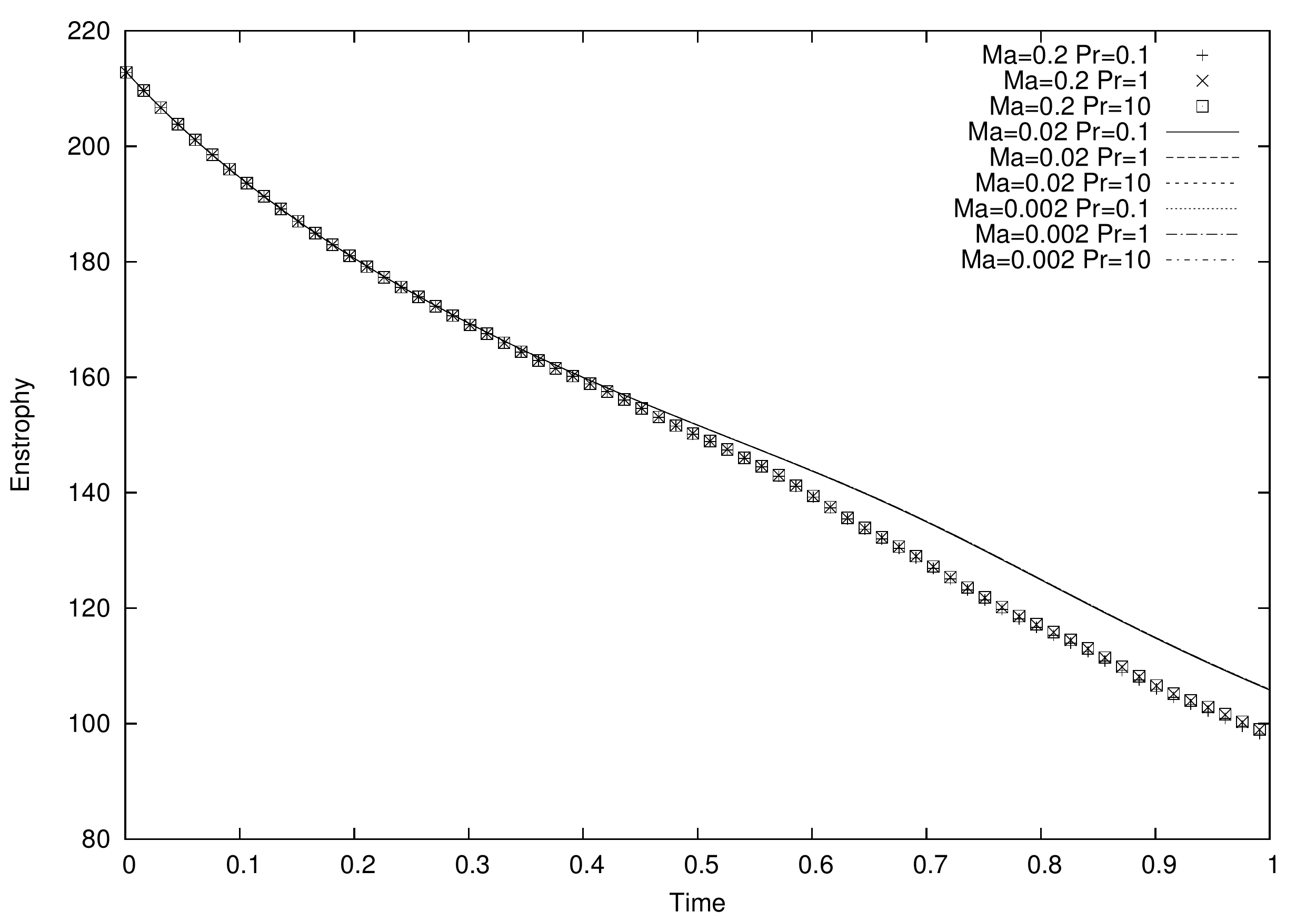}
\caption{Enstrophy decay for different Mach and Prandtl numbers.}
\label{compaenstrophy}
\end{figure}

\begin{figure}[h]
\includegraphics[angle=-90,width=\textwidth]{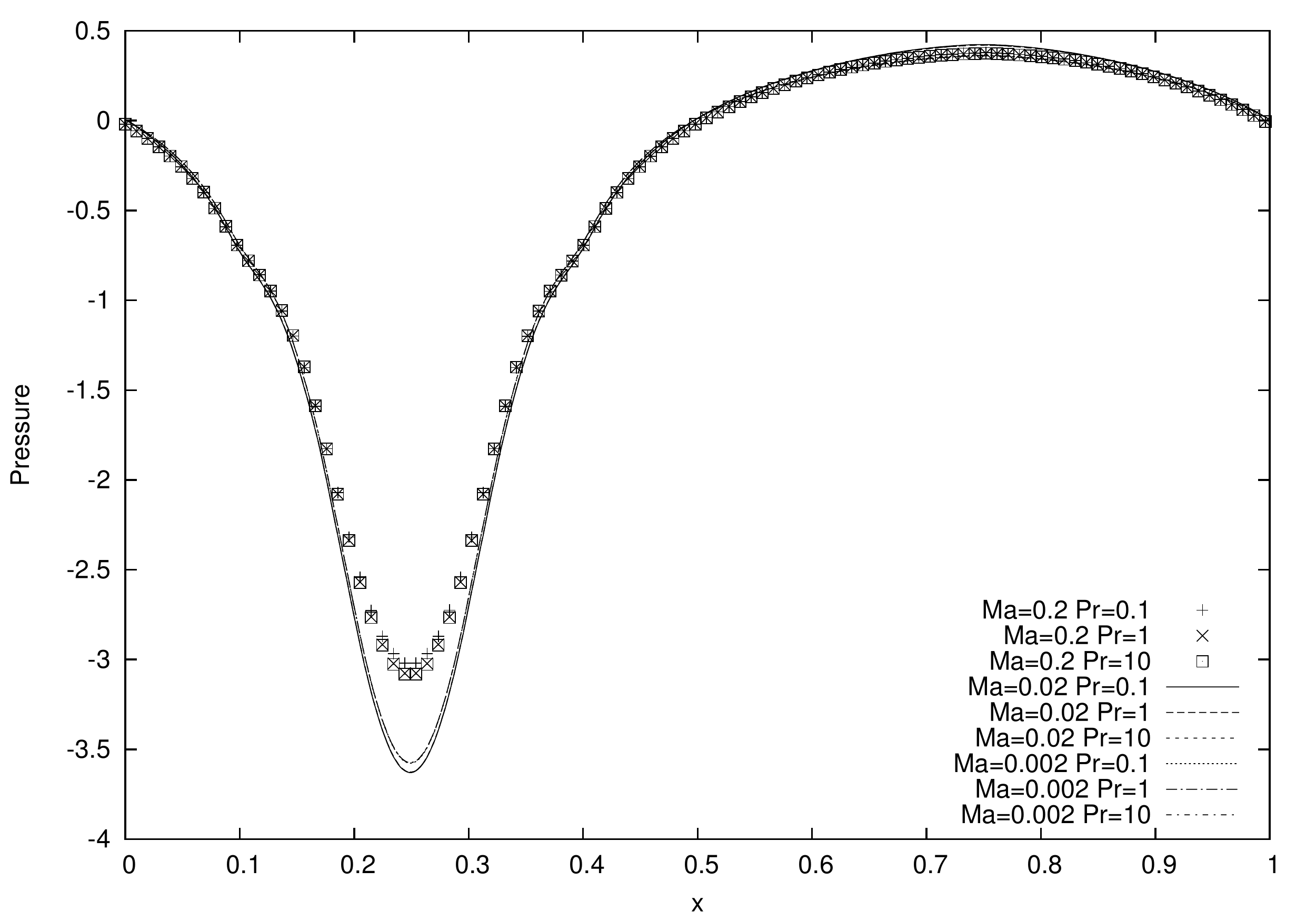}
\includegraphics[angle=-90,width=\textwidth]{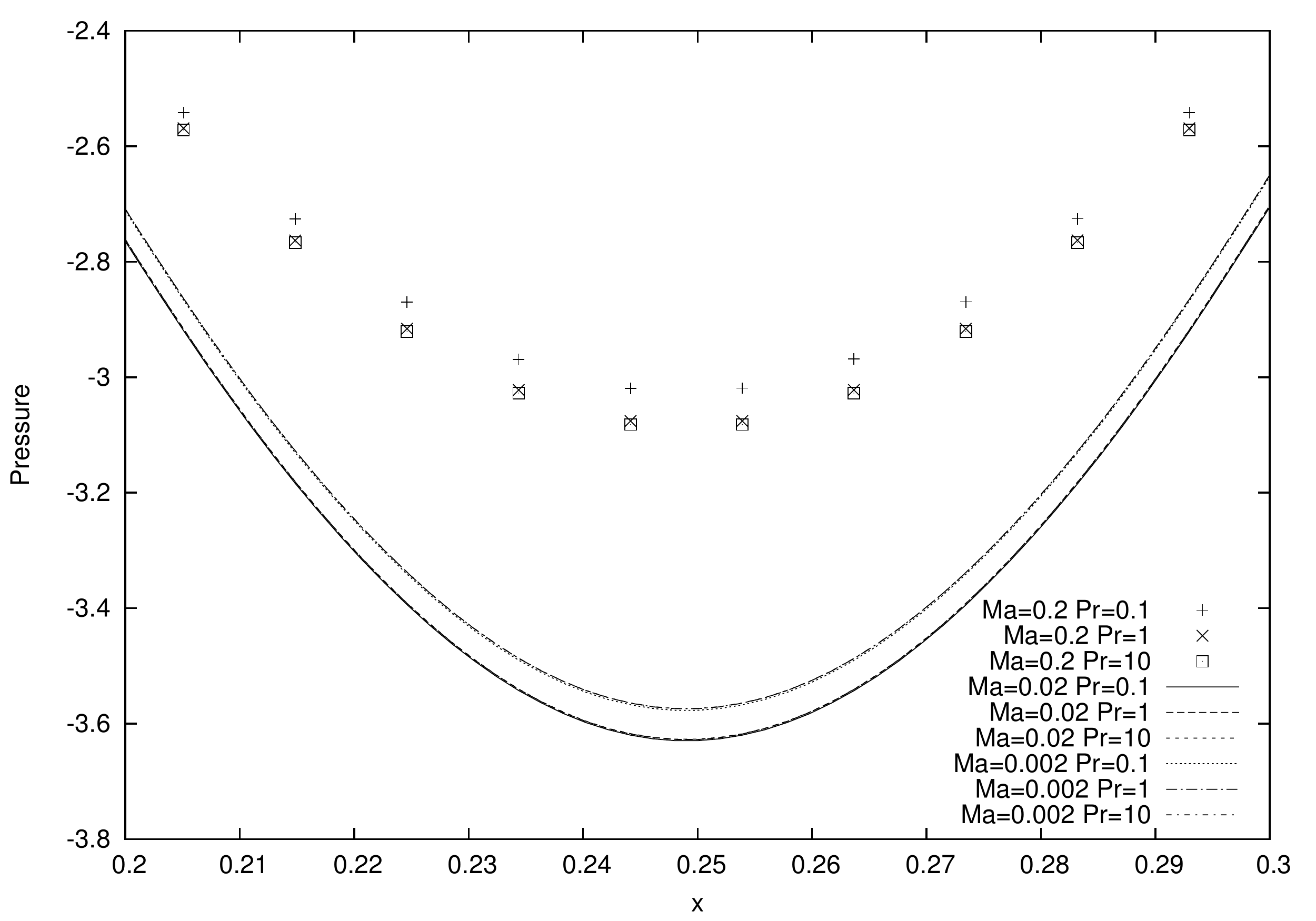}
\caption{Spatial evolution of pressure in function of x at y=0.25 and t=1 (top: whole profile, bottom: zoom around the peak x$\in[0.2,0.3]$). Effects of the Prandtl and Mach numbers.}
\label{compapMaPr}
\end{figure}

\begin{figure}[h]
\includegraphics[angle=-90,width=\textwidth]{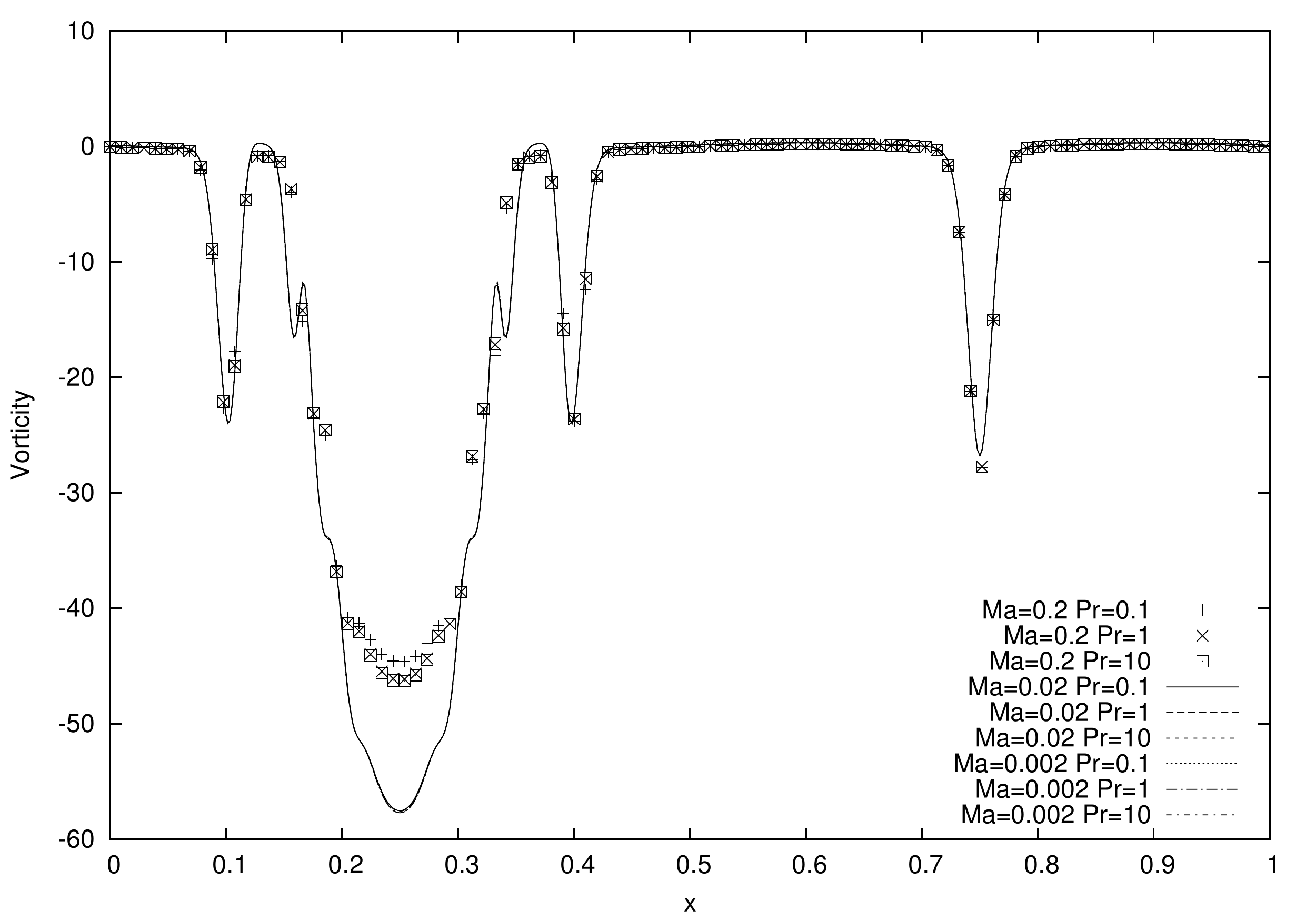}
\includegraphics[angle=-90,width=\textwidth]{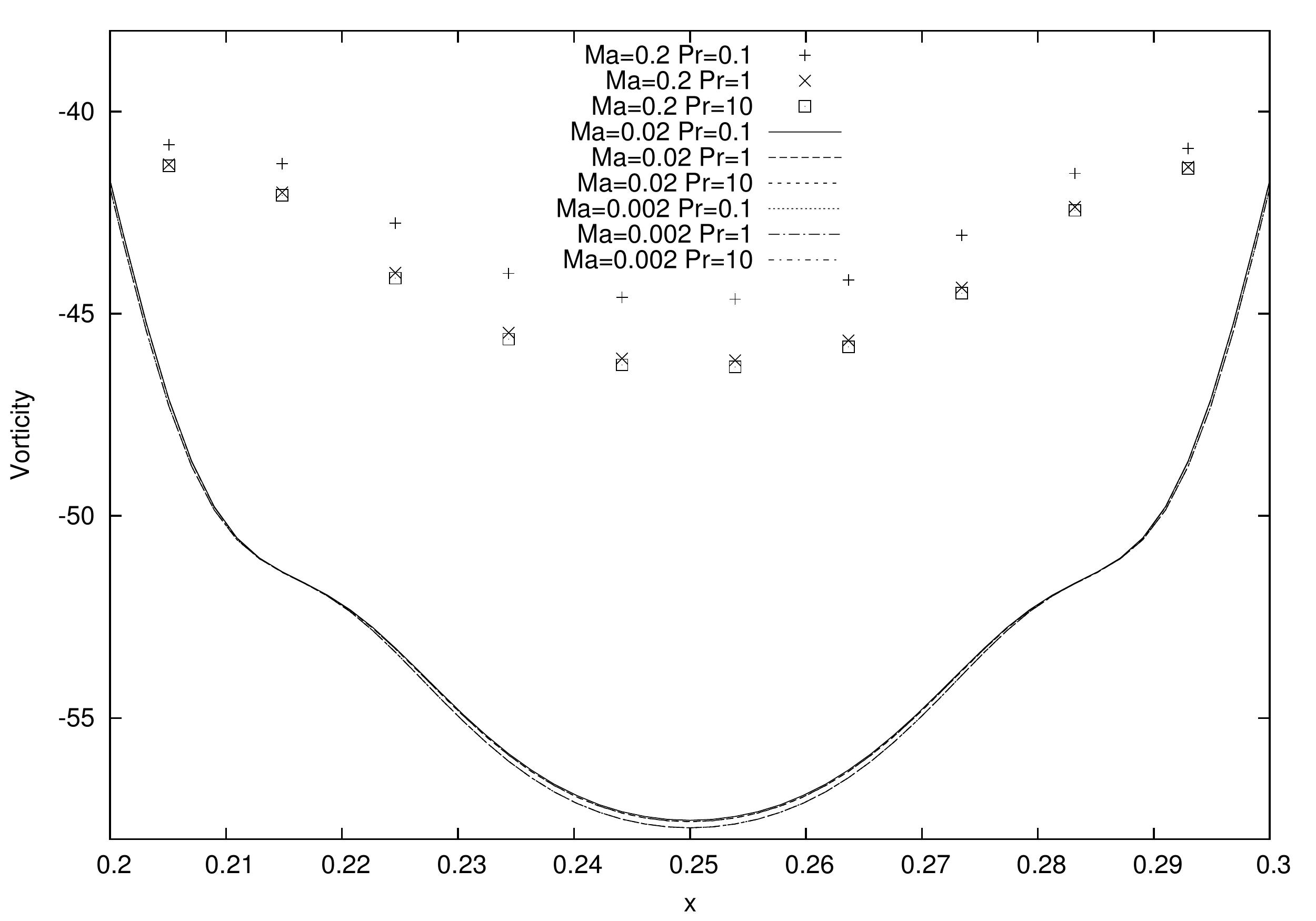}
\caption{Spatial evolution of vorticity in function of x at y=0.25 and t=1 (top: whole profile, bottom: zoom around the peak x$\in[0.2,0.3]$). Effects of the Prandtl and Mach numbers.}
\label{compavorticityMaPr}
\end{figure}

\artrois{Figures~\ref{compakinetic},~\ref{compaenstrophy},~\ref{compapMaPr} and~\ref{compavorticityMaPr} show the results obtained for all the pairs $(Ma, Pr)$ with $Ma\in \{0.2,~0.02~and~0.002\}$ and $Pr\in \{0.1,~1~and~10\}$}. Figures~\ref{compakinetic} and~\ref{compaenstrophy} show that a Mach number $Ma=0.2$ is too big and modifies integral quantities such as kinetic energy and enstrophy. In contrast, the Prandtl number seems to have no effect on these integral quantities. The spatial evolution of pressure and vorticity presented in figures~\ref{compapMaPr} and~\ref{compavorticityMaPr} confirms that $Ma=0.2$ modifies the results and that the Prandtl number has almost no effect. 

\begin{figure}[h]
\includegraphics[angle=-90,width=\textwidth]{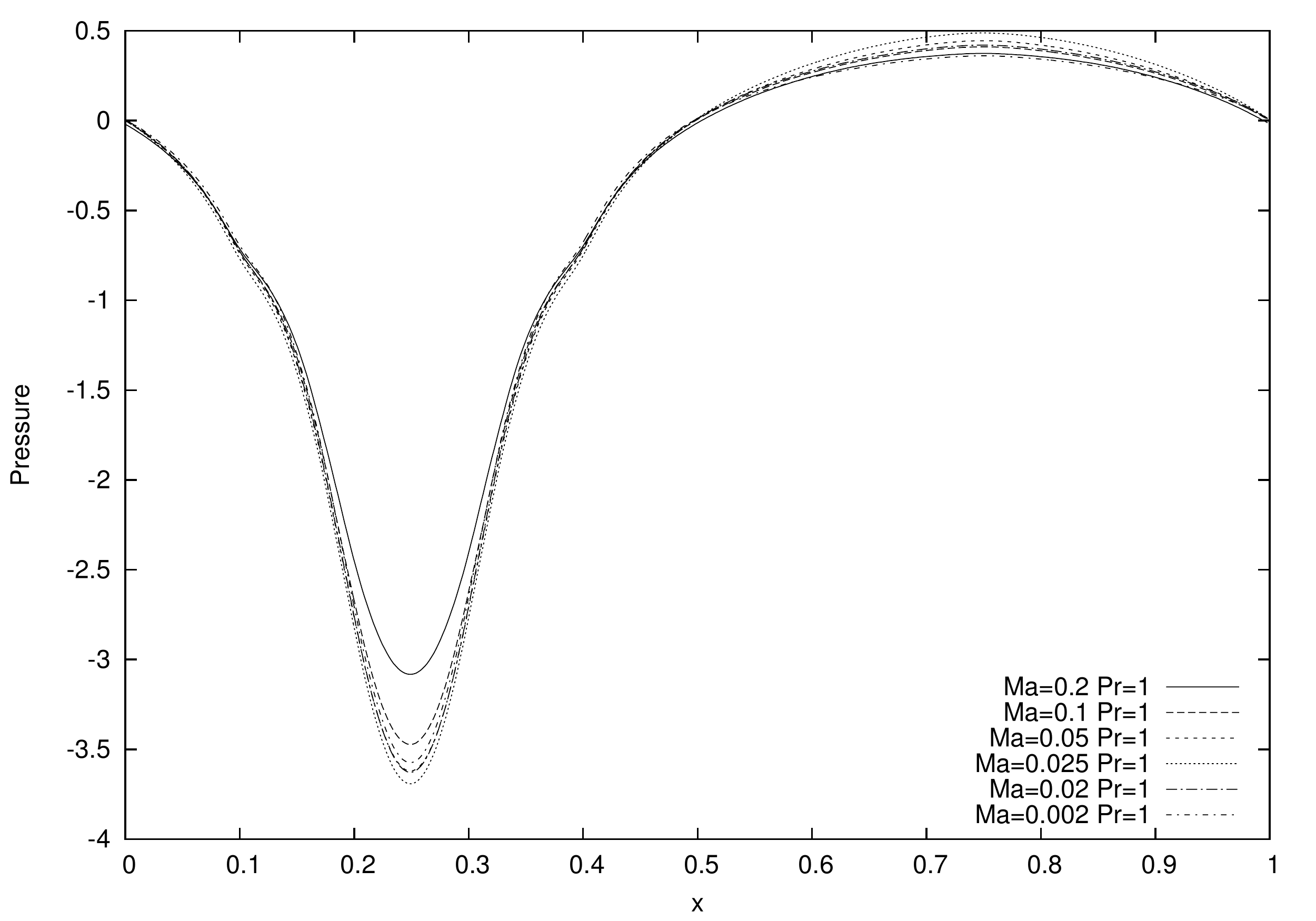}
\caption{Spatial evolution of pressure in function of x at y=0.25 and t=1. Effect of the Mach number.}
\label{compapMa}
\end{figure}

\begin{figure}[h]
\includegraphics[angle=-90,width=\textwidth]{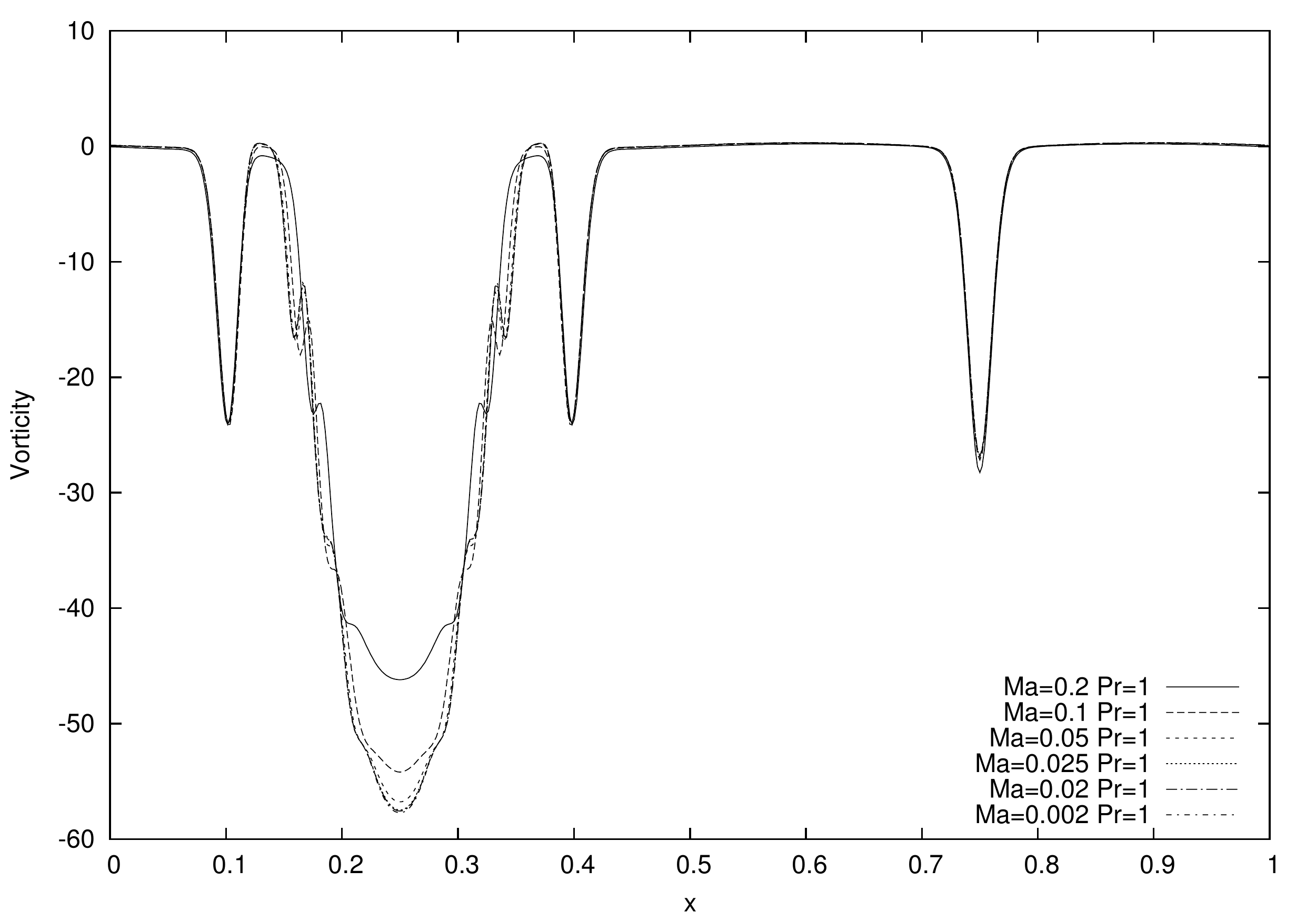}
\caption{Spatial evolution of vorticity in function of x at y=0.25 and t=1. Effect of the Mach number.}
\label{compavorticityMa}
\end{figure}

The same profiles with more values of the Mach number and $Pr=1$ are plotted in figures~\ref{compapMa} and~\ref{compavorticityMa}. Very low Mach numbers do not imply accuracy problems but they imply very small time-step. Indeed, due to acoustic, the time scale of GPE, $t_{GPE}$, is related to that of INS equations, $t_{INS}$; $ t_{GPE} = Ma t_{INS}$.   A Mach number equal to 0.05 -value typically used in lattice Boltzmann simulations- offers a good compromize between the accuracy of the results and the stability time step. This value is coherent with the order of magnitude proposed for KRLNS by~\cite{karlin:2006}. For GPE, this value has to be confirm in other configurations.


\begin{figure}[h]
\includegraphics[angle=-90,width=\textwidth]{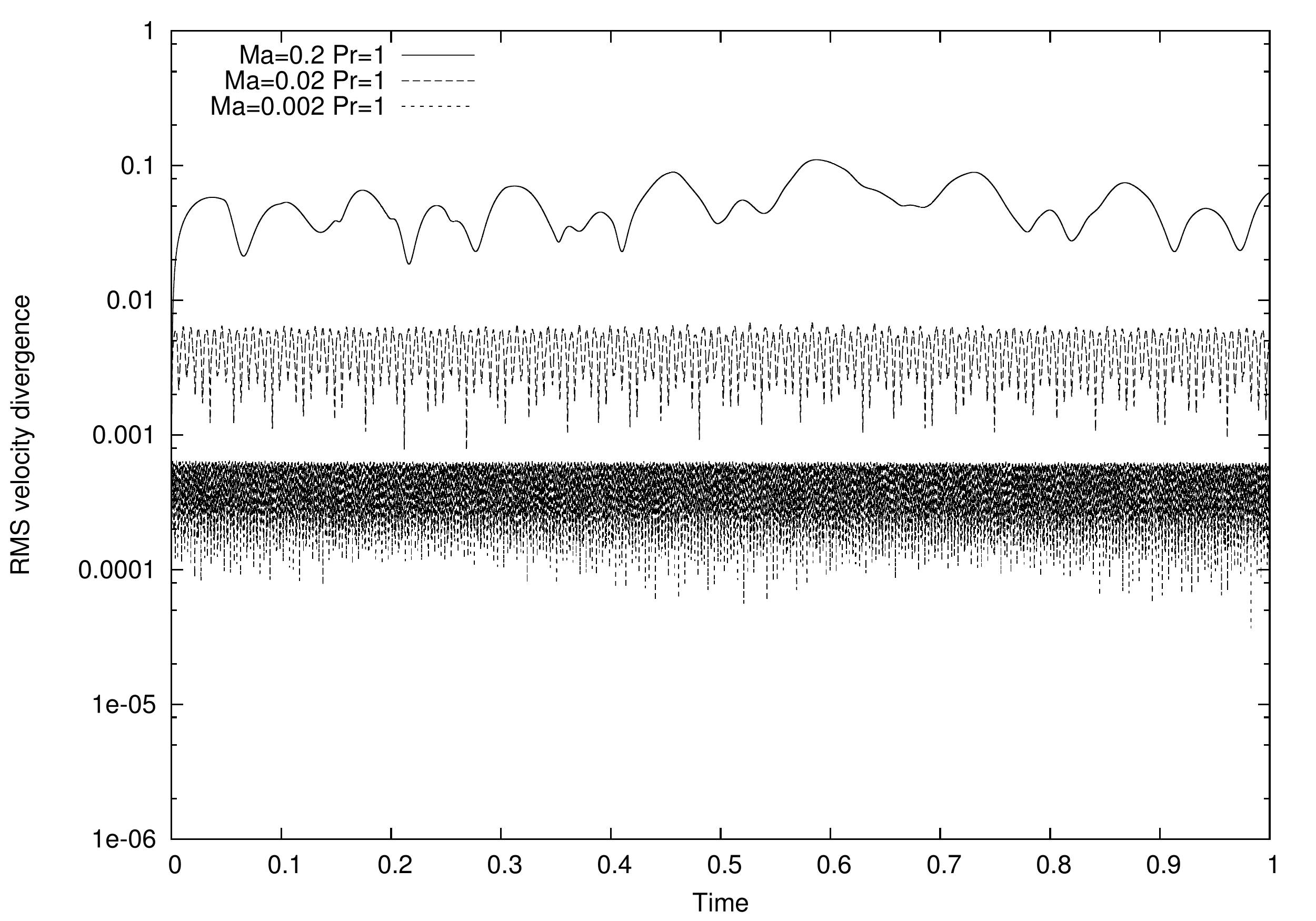}
\includegraphics[angle=-90,width=\textwidth]{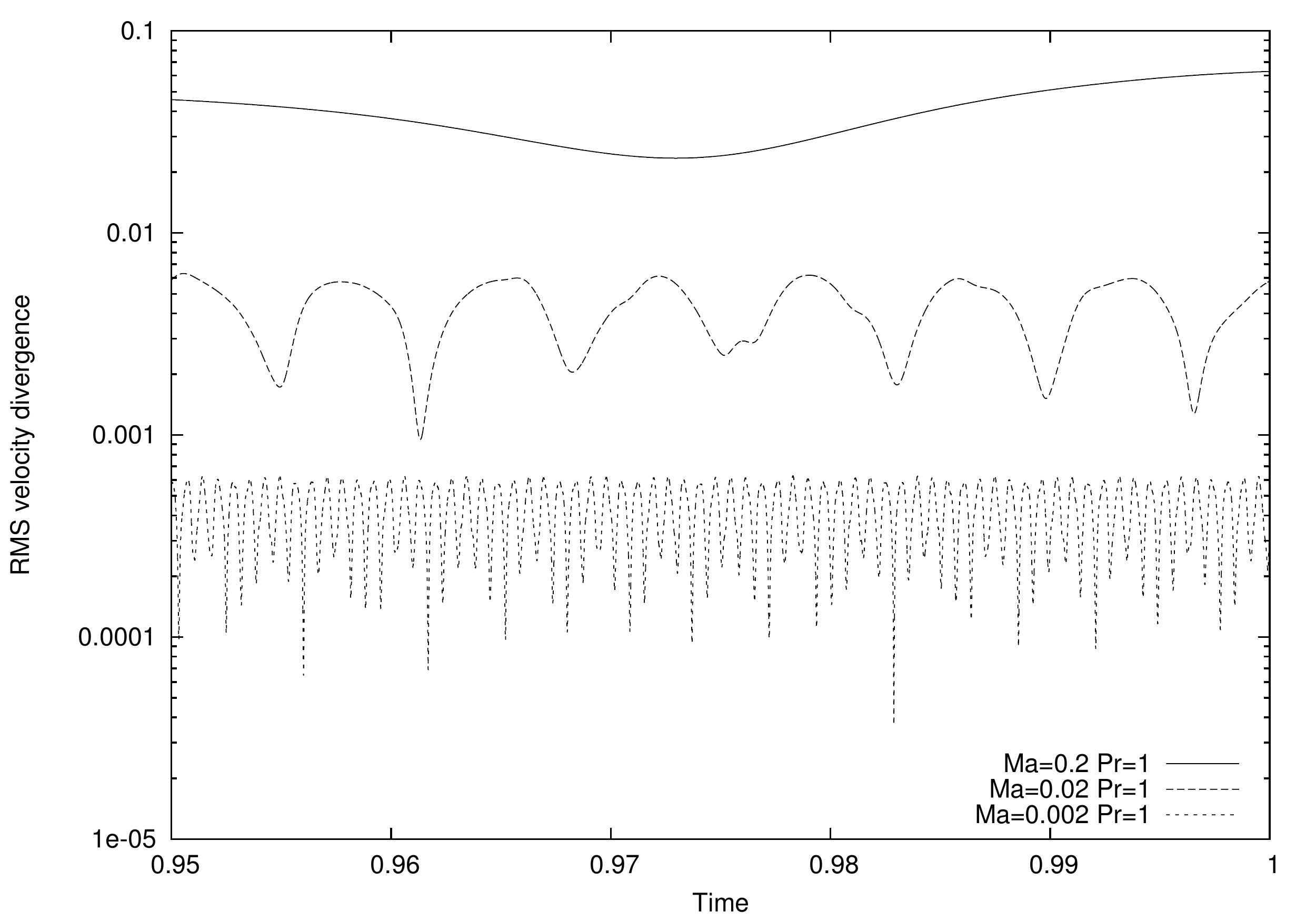}
\caption{Time evolution of root mean square velocity divergence (top: whole time of the simulation, bottom: zoom on the last 0.05s). Effect of the Mach number.}
\label{compadivuMa}
\end{figure}

The rms velocity divergence is defined by
\BE
(\D_\alpha u_\alpha)_{rms}=\sqrt{\int_V (\D_\alpha u_\alpha)^2dV}
\label{divurms}
\EE
It is worth noting that the mean velocity divergence $\int_V (\D_\alpha u_\alpha)dV$ is exactly equal to zero due to periodic conditions and the used staggered grid system. Figure~\ref{compadivuMa} shows that the maximum of the rms velocity divergence is proportional to the Mach number. The proportionality constant seems to be equal to 0.3. This linear scaling and this proportionality constant have also been found for KRLNS~\cite{karlin:2006}. A Fourier transform of the temporal signal gives that the nondimensionalized period is also proportional to the Mach number ($T=7.10^{-3}$ for $Ma=0.02$ and $T=7.10^{-4}$ for $Ma=0.002$\footnote{For Ma=0.2, the signal is not enough periodic to isolate one frequency.}). The period corresponds to the plane wave associated to the thickness of the shear layer. Indeed, noting $\delta_{sl}$ the nondimensionalized shear layer thickness, one gets $\delta_{sl}\approx\F{T}{2\pi Ma}=0.056$.

\begin{figure}[h]
\includegraphics[angle=-90,width=\textwidth]{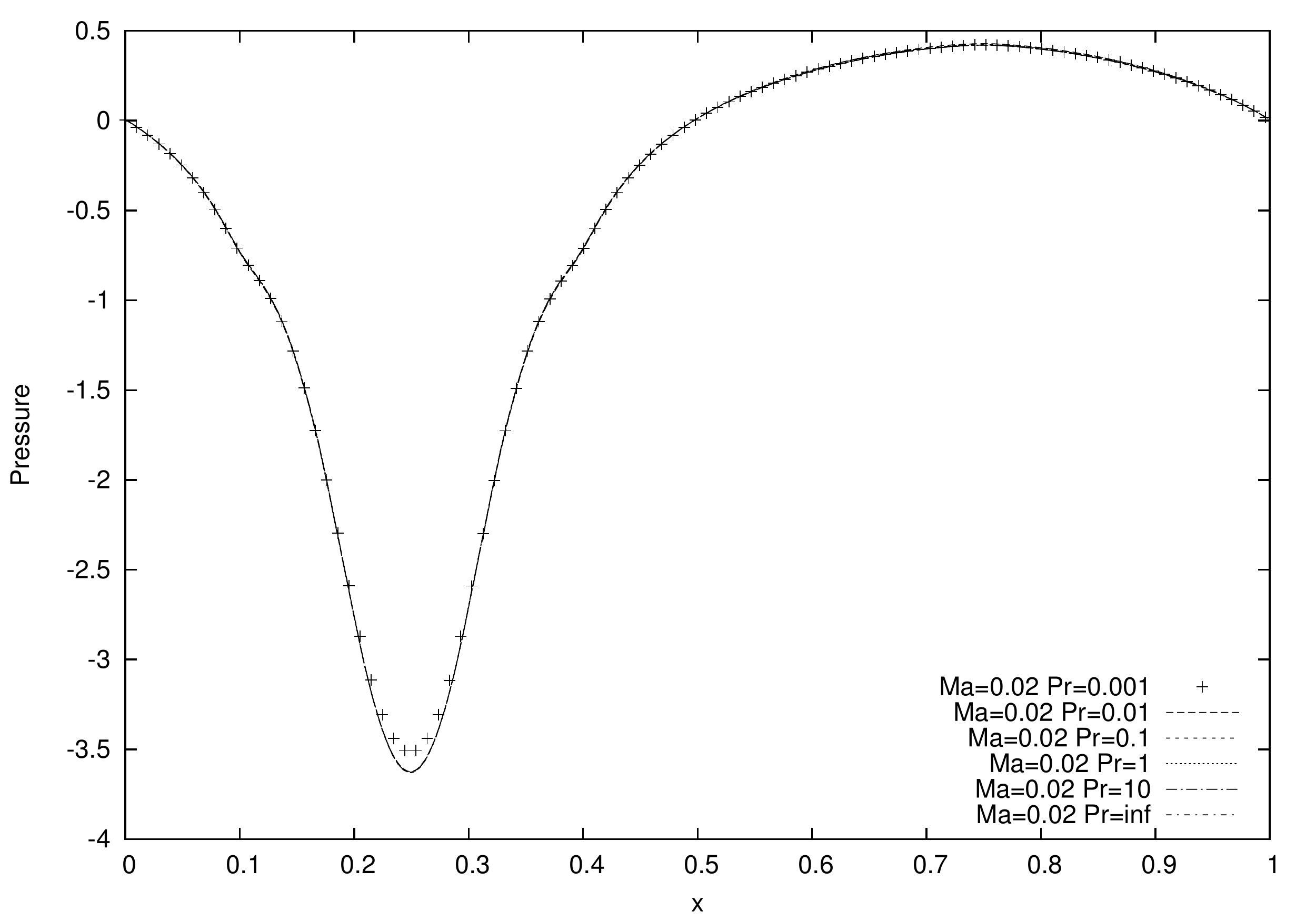}
\caption{Spatial evolution of pressure in function of x at y=0.25 and t=1. Effect of the Prandtl number.}
\label{compapPr}
\end{figure}

\begin{figure}[h]
\includegraphics[angle=-90,width=\textwidth]{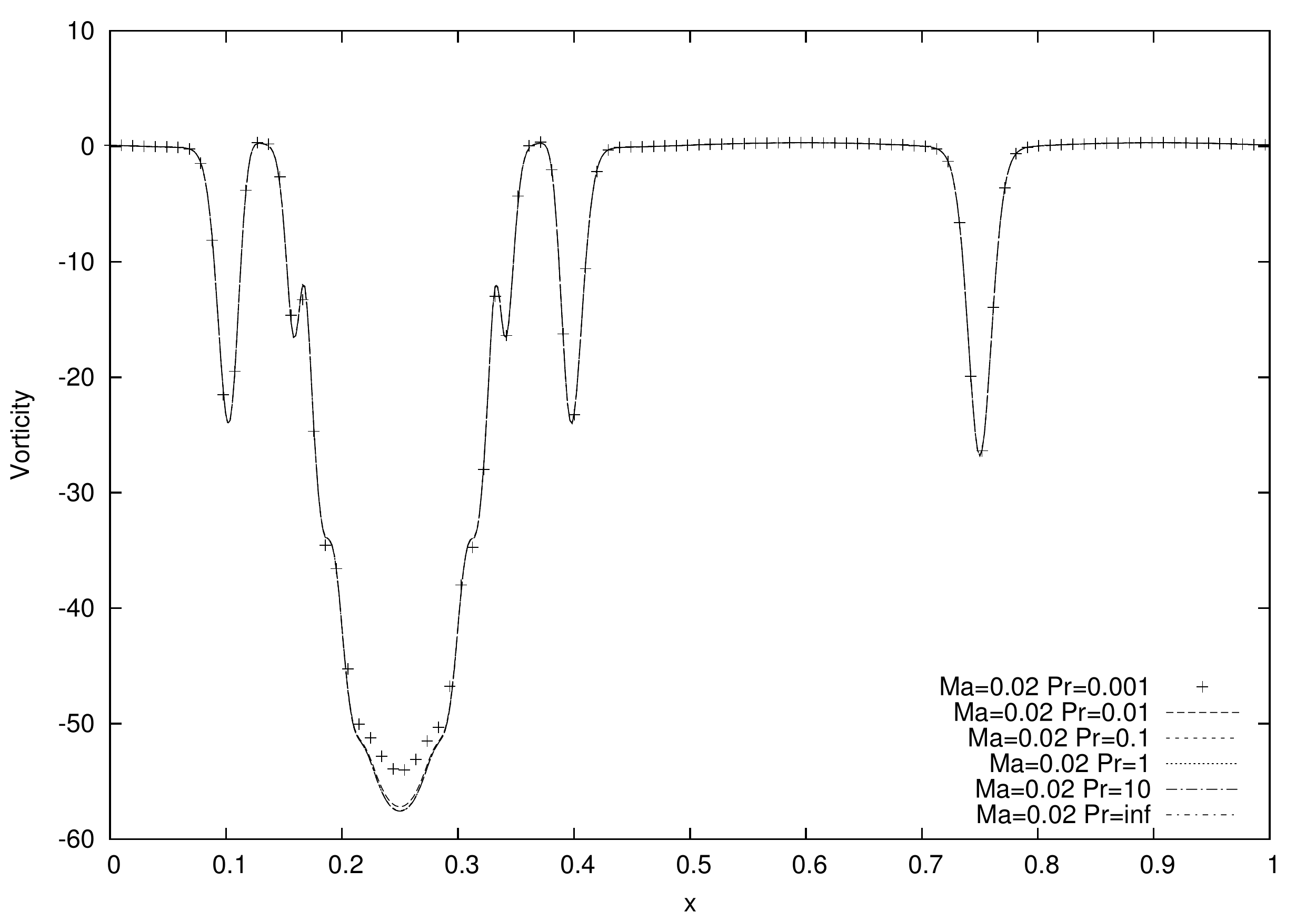}
\caption{Spatial evolution of vorticity in function of x at y=0.25 and t=1. Effect of the Prandtl number.}
\label{compavorticityPr}
\end{figure}

\begin{figure} 
\centering 

\begin{subfigure}{0.4\textwidth}
\centering 
\adjincludegraphics[width=\textwidth]{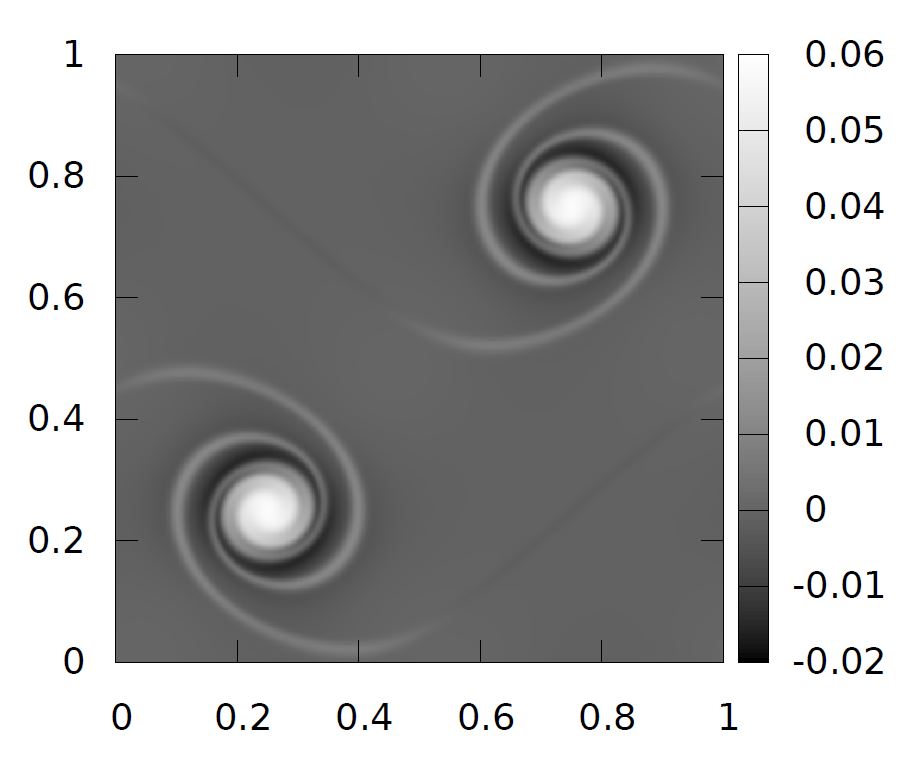}
\caption{Pr=0.001}
\end{subfigure}
\hspace*{2cm}
\begin{subfigure}{0.4\textwidth}
\centering 
\adjincludegraphics[width=\textwidth]{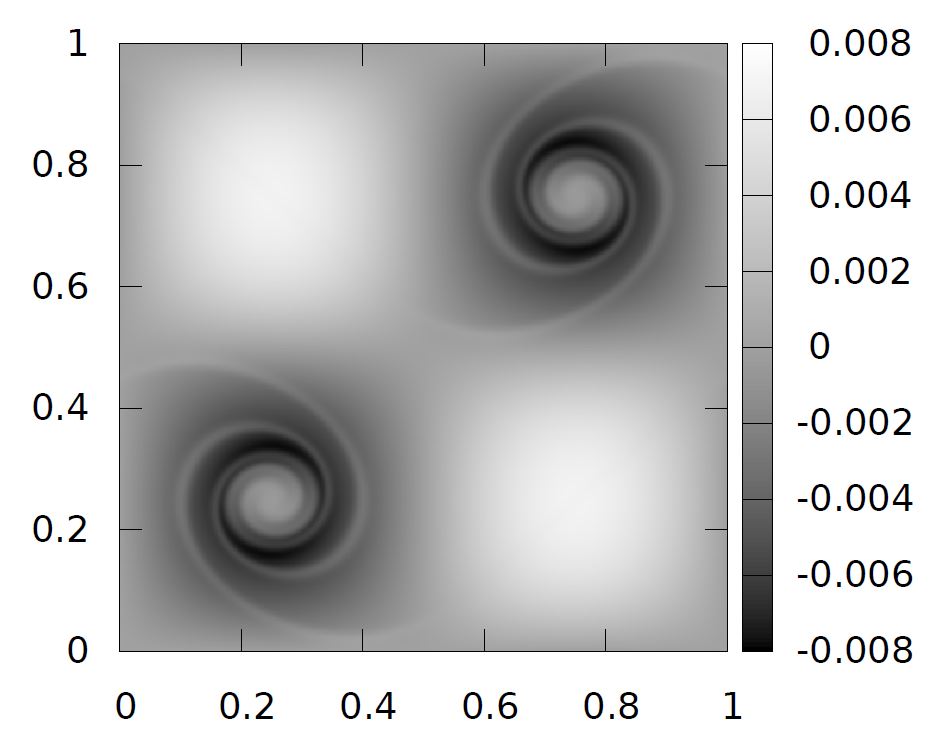}
\caption{Pr=0.01}
\end{subfigure}

\begin{subfigure}{0.4\textwidth}
\adjincludegraphics[width=\textwidth]{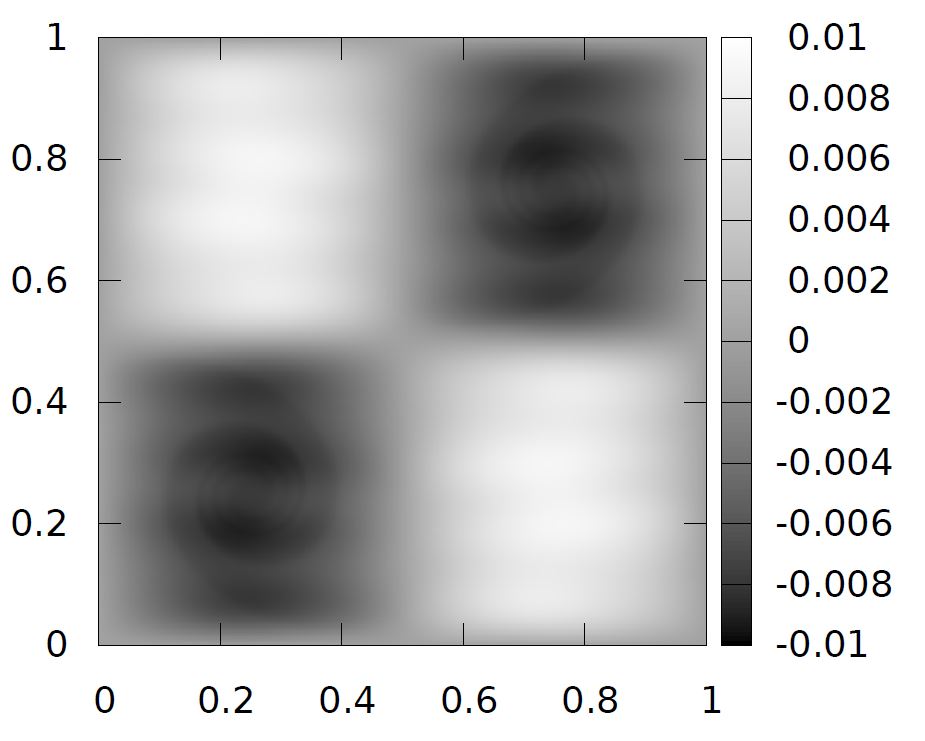}
\caption{Pr=0.1}
\end{subfigure}
\hspace*{2cm}
\begin{subfigure}{0.4\textwidth}
\adjincludegraphics[width=\textwidth]{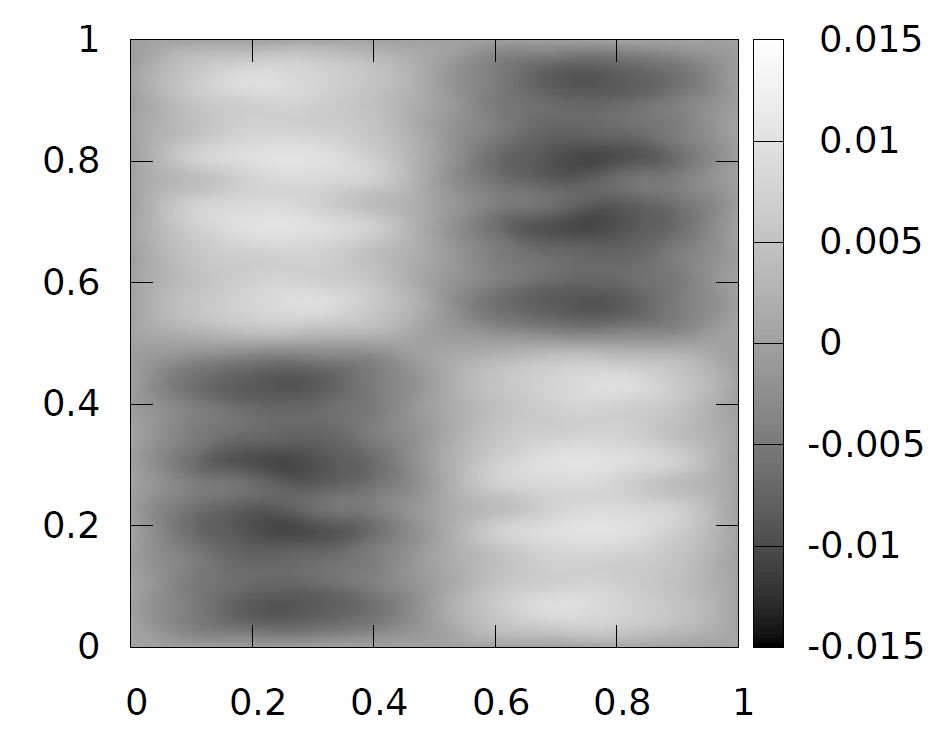}
\caption{Pr=1}
\end{subfigure}

\begin{subfigure}{0.4\textwidth}
\adjincludegraphics[width=\textwidth]{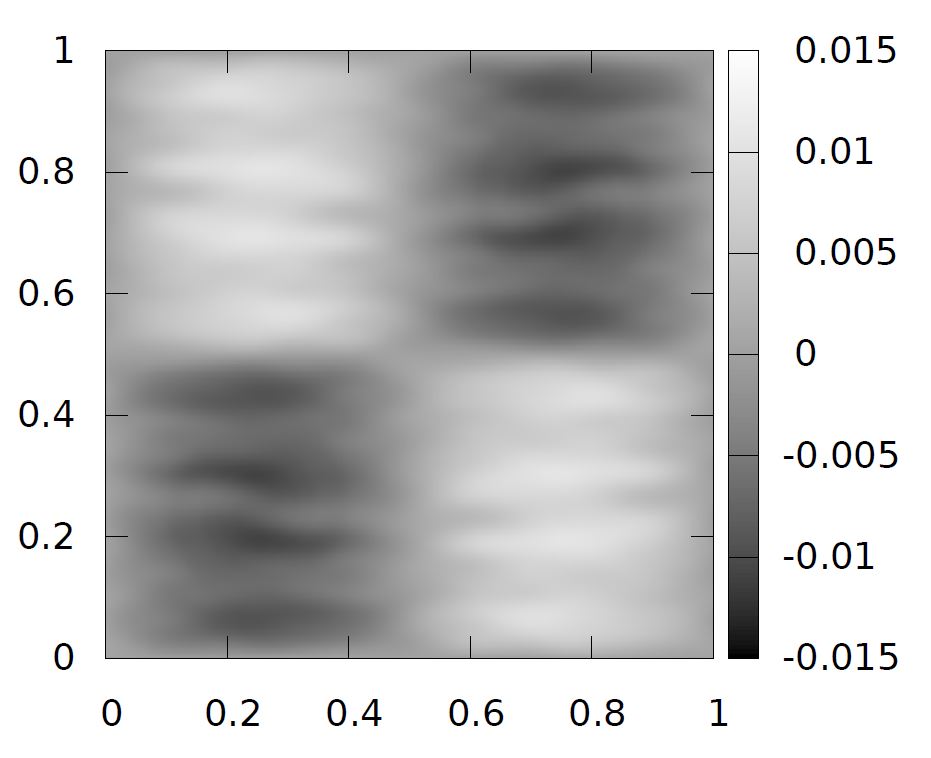}
\caption{Pr=10}
\end{subfigure}
\hspace*{2cm}
\begin{subfigure}{0.4\textwidth}
\adjincludegraphics[width=\textwidth]{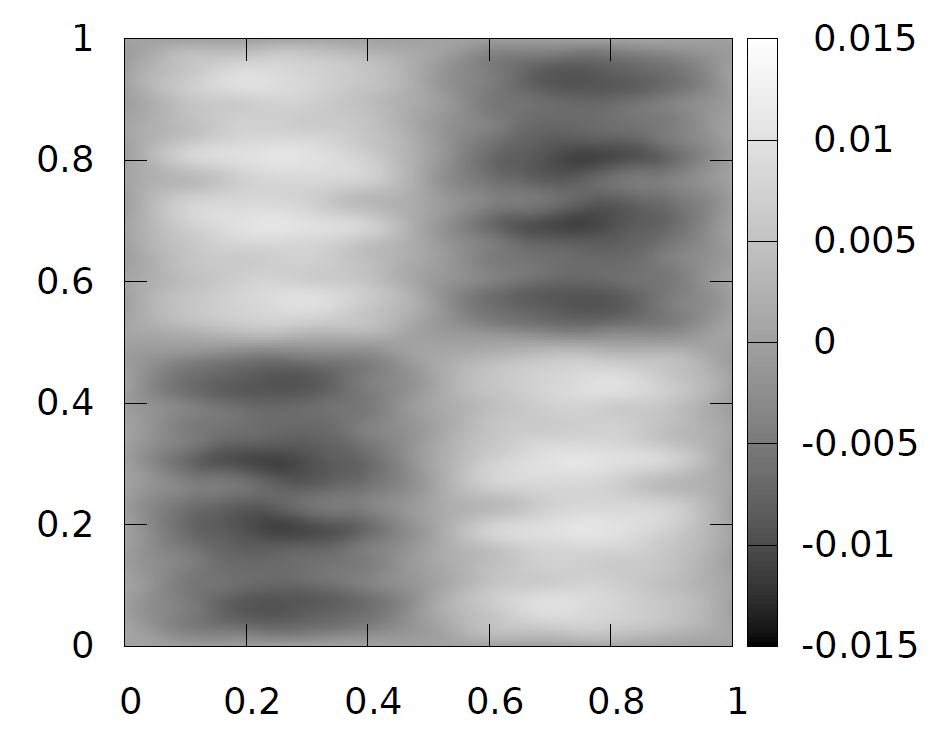}
\caption{$Pr=+\infty$}
\end{subfigure}

\caption{Velocity divergence at t=1 for Ma=0.02 and for different Prandtl numbers.}
\label{divuPr}
\end{figure}

\begin{figure} 
\centering 
\begin{subfigure}{0.4\textwidth}
\adjincludegraphics[width=\textwidth]{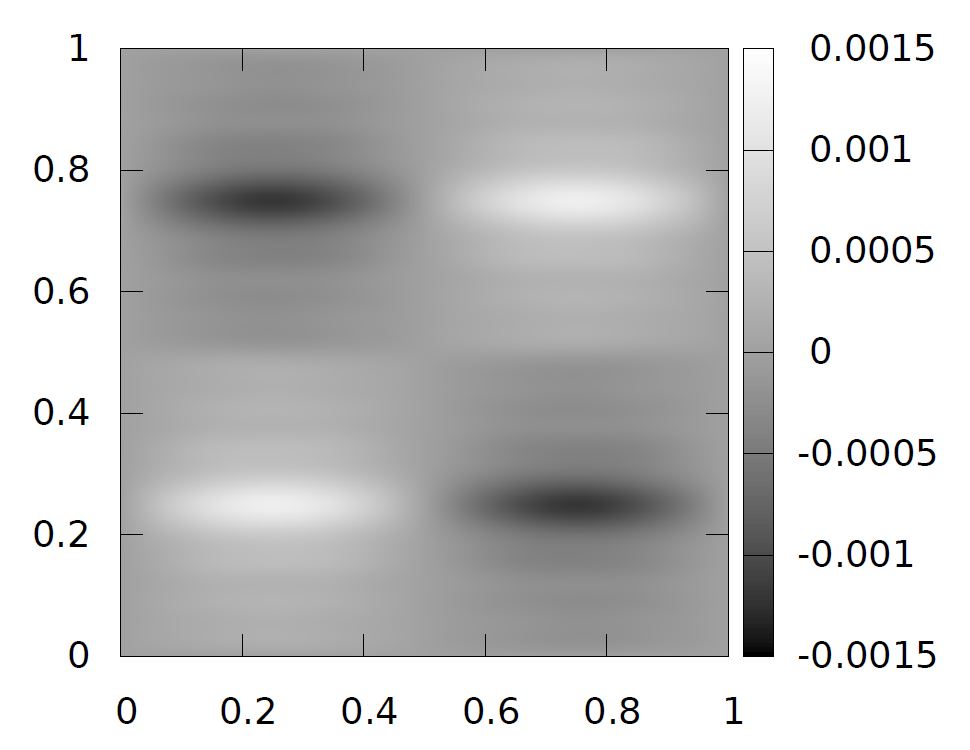}
\caption{\artrois{Ma=0.002}}
\end{subfigure}
\hspace*{2cm}
\begin{subfigure}{0.4\textwidth}
\centering 
\adjincludegraphics[width=\textwidth]{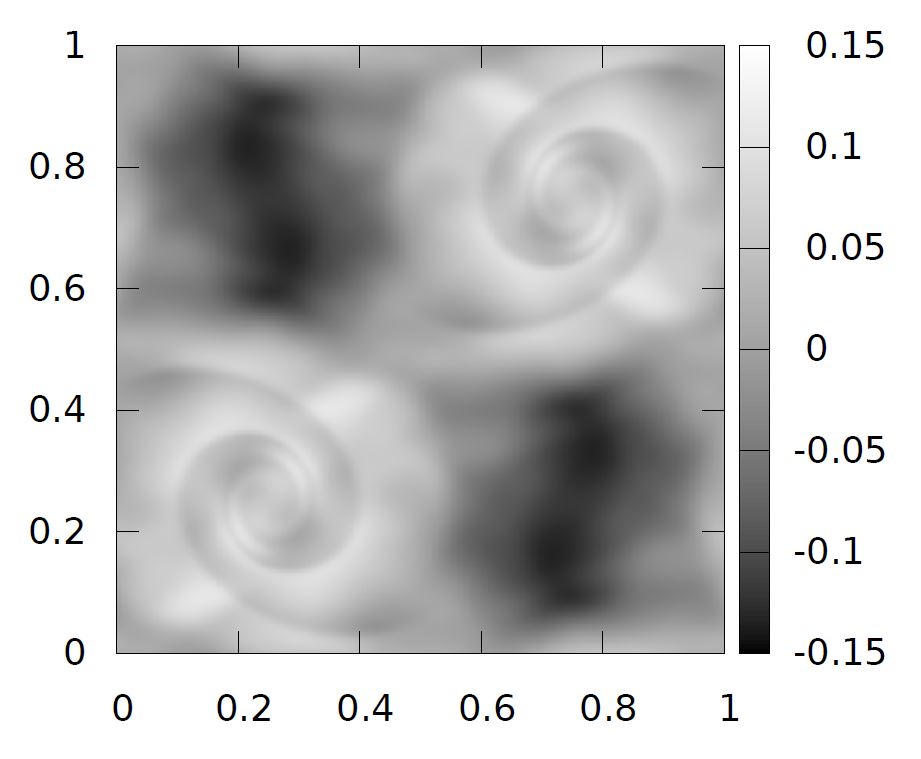}
\caption{\artrois{Ma=0.2}}
\end{subfigure}
\caption{Velocity divergence at t=1 for Pr=1 and for different Mach numbers.}
\label{divuMa}
\end{figure}

Figures~\ref{compapPr} and~\ref{compavorticityPr} seem to show that the Prandtl number has almost no effect on velocity and pressure. However, figure~\ref{divuPr} indicates that the Prandtl number impacts significantly the velocity divergence. Indeed, according to the Prandtl number, we can distinguish the acoustic waves or/and the shear layer vortex in the velocity divergence field. For a Prandtl number smaller or equal to 0.001, acoustic waves are completely damped at $t=1$ and only shear layer vortex are observable. For a Prandtl number between 0.01 and 0.1, both acoustic waves and shear layer vortex are present.  For a Prandtl number greater or equal to one, only acoustic waves are visible. Furthermore, it is worth noting that, even if a small Prandtl number conducts to acoustic wave damping, a smaller Prandtl number does not imply a smaller velocity divergence. These observations are coherent with the acoustic plane wave solution of GPE equations. In one dimension with the acoustic approximation, GPE equations are
\begin{subequations}
\begin{eqnarray}
\D_tu+\D_xP&=&\F{1}{Re}\D_x\D_xu\\
\D_tP+\F{1}{Ma^2}\D_xu&=&\F{1}{RePr}\D_x\D_xP
\end{eqnarray}
\end{subequations}
The acoustic plane wave solution is given by
\begin{subequations}
\begin{eqnarray}
u&=&\left(\F{1-Pr}{2RePr}ksin(\omega t-kx)+\F{\omega}{k}cos(\omega t-kx)\right)Ma^2e^{-at}\\
P&=&cos(\omega t-kx)e^{-at}\\
\omega^2&=&k^2\left(\F{1}{Ma^2}-\F{1}{4Re^2}\left(1-\F{1}{Pr}\right)^2k^2\right)\\
a&=&\F{1+Pr}{2RePr}k^2\label{dampcoeff}
\end{eqnarray}
\end{subequations}
For $Pr=1$, this solution gives the classical relation $k^2=\F{\omega^2}{c^2}$ (in dimensional variables and with $c$ the speed of sound). The expression of the damping coefficient is coherent with the fact that both viscosity and thermal conductivity are involved in acoustic wave attenuation~\cite{landau:1987}. The smaller the Prandtl number, the higher the damping coefficient ($\displaystyle{\lim_{Pr \to 0} a=+\infty}$). \artrois{One can compare the effect of the Mach and Prandtl numbers on the velocity divergence in figures~\ref{divuPr} and~\ref{divuMa}. A small Mach number allows the decrease of both acoustic waves and shear layer vortex in the velocity divergence field. A small Prandtl number allows the damping of acoustic waves but not of the shear layer vortex. A smaller Mach number implies a smaller velocity divergence. It is not the case for the Prandtl number.}

\section{Conclusion}
The proposed general pressure equation (GPE) allows to solve general, time-accurate, incompressible Navier-Stokes flows. GPE  should be considered as an analog of the Lattice Boltzmann Method. The velocity divergence is not equal to zero but it can be limited by an arbitrary maximum. General pressure equation (GPE), artificial compressibility method (ACM) and \edac{entropically damped form of artificial compressibility (EDAC)} have been applied for simulations of Taylor-Green vortices, traveling wave and doubly periodic shear layers in order to investigate their accuracy, efficiency and the capability to capture the correct transient behavior. The proposed discretization uses a staggered grid system, second order centered schemes in space and a third order Runge-Kutta scheme in time without subiteration. The solutions obtained by GPE, ACM \edac{and EDAC} are in excellent agreement with analytical solutions or with that of a classical finite volume method using a Poisson equation. It is found that the GPE, ACM \edac{and EDAC} approaches can capture the correct transient behavior without subiteration, and keep the divergence fluctuation at smaller level by giving an appropriately low Mach number. \edac{GPE L2 norm of errors are smaller than those of ACM and EDAC}. For Taylor-Green vortices, the better convergence rate of GPE is explained by the fact that this flow is analytical solutions of GPE but not of ACM \edac{and EDAC}. Consequently,  the additional diffusion term of GPE \edac{compared to ACM} seems physically crucial \edac{and the additional pressure advection term of EDAC compared to GPE seems to be useless}. The one-dimensional acoustic plane wave analytical solution of GPE gives the acoustic damping coefficient. This coefficient is inversely proportional to the Prandtl number (for small Prandtl number). Numerically, it is shown that a small Prandtl number allows to damp acoustic waves.

\bibliographystyle{elsarticle-num}
\bibliography{biblio}

\end{document}